%% file: StopCharmRun2Paper.tex
\pdfoutput=1
\newcommand*{\ATLASLATEXPATH}{latex/}
\documentclass[UKenglish, cernpreprint, texlive=2016]{\ATLASLATEXPATH atlasdoc}
\usepackage{\ATLASLATEXPATH atlaspackage}

\usepackage{\ATLASLATEXPATH atlasbiblatex} 

\usepackage{\ATLASLATEXPATH atlasphysics}

\usepackage{array}
\newcolumntype{P}[1]{>{\centering\arraybackslash}p{#1}}

\graphicspath{{logos/}{figures/}}

\usepackage{amsbsy}
\usepackage{multirow}
\usepackage{StopCharmRun2Paper-defs}
\usepackage{latex/atlasbsm}

\input{StopCharmRun2Paper-metadata}
\hypersetup{pdftitle={ATLAS document},pdfauthor={The ATLAS Collaboration}}

\begin{document}

\maketitle

%\tableofcontents

%-------------------------------------------------------------------------------
\section{Introduction}
\label{sec:intro}
%-------------------------------------------------------------------------------

\input{latex/sections/Introduction.tex}

%-------------------------------------------------------------------------------
\section{ATLAS detector}
\label{sec:detector}
%-------------------------------------------------------------------------------

\input{latex/sections/Detector.tex}

%-------------------------------------------------------------------------------
\section{Data and simulated event samples}
\label{sec:dataAndMC}
%-------------------------------------------------------------------------------

\input{latex/sections/DataAndMC.tex}

%-------------------------------------------------------------------------------
\section{Event Reconstruction}
\label{sec:reco}
%-------------------------------------------------------------------------------

\input{latex/sections/EventReco.tex}
\label{sec:eventReco}

%-------------------------------------------------------------------------------
\section{Event Selection}
\label{sec:selection}
%-------------------------------------------------------------------------------

\input{latex/sections/EventSelection.tex}

%-------------------------------------------------------------------------------
\section{Background estimation}
\label{sec:background}
%-------------------------------------------------------------------------------

\input{latex/sections/BackgroundEstimation.tex}

%-------------------------------------------------------------------------------
\section{Systematic Uncertainties}
\label{sec:uncert}
%-------------------------------------------------------------------------------

\input{latex/sections/SystematicsUncertainties.tex}

%-------------------------------------------------------------------------------
\section{Results}
\label{sec:result}
%-------------------------------------------------------------------------------

\input{latex/sections/Results.tex}

% All figures and tables should appear before the summary and conclusion.
% The package placeins provides the macro \FloatBarrier to achieve this.
\FloatBarrier

%-------------------------------------------------------------------------------
\section{Conclusion}
\label{sec:conclusion}
%-------------------------------------------------------------------------------

\input{latex/sections/Conclusion.tex}

%-------------------------------------------------------------------------------
\section*{Acknowledgements}
%-------------------------------------------------------------------------------

\input{acknowledgements/Acknowledgements}

%
%The \texttt{atlaslatex} package contains the acknowledgements that were valid 
%at the time of the release you are using.
%These can be found in the \texttt{acknowledgements} subdirectory.
%When your ATLAS paper or PUB/CONF note is ready to be published,
%download the latest set of acknowledgements from:\\
%\url{https://twiki.cern.ch/twiki/bin/view/AtlasProtected/PubComAcknowledgements}

%%-------------------------------------------------------------------------------
%\clearpage
%\appendix
%\part*{Appendix}
%\addcontentsline{toc}{part}{Appendix}
%%-------------------------------------------------------------------------------
%
%In a paper, an appendix is used for technical details that would otherwise disturb the flow of the paper.
%Such an appendix should be printed before the Bibliography.

%-------------------------------------------------------------------------------
% If you use biblatex and either biber or bibtex to process the bibliography
% just say \printbibliography here
\printbibliography
% If you want to use the traditional BibTeX you need to use the syntax below.
%\bibliographystyle{bib/bst/atlasBibStyleWoTitle}
%\bibliography{StopCharmRun2Paper,bib/ATLAS,bib/CMS,bib/ConfNotes,bib/PubNotes}
%-------------------------------------------------------------------------------

\clearpage \input{atlas_authlist}

%-------------------------------------------------------------------------------
% Auxiliary material - comment out the following line if you do not have any
%\include{StopCharmRun2Paper-auxmat}
%-------------------------------------------------------------------------------

%-------------------------------------------------------------------------------
% Extra tables etc. for HepData - comment in the following line if you have any
% \include{StopCharmRun2Paper-hepdata}
%-------------------------------------------------------------------------------

\end{document}

%% file: StopCharmRun2Paper-metadata.tex
\AtlasTitle{Search for supersymmetry in final states with charm jets and missing transverse momentum in \SI{13}{\TeV} \pp{} collisions with the ATLAS detector}

\AtlasAbstract{%
A search for supersymmetric partners of top quarks decaying as $\stopone\to c\ninoone$ and supersymmetric partners of charm quarks decaying as $\scharmone\to c\ninoone$, where $\ninoone$ is the lightest neutralino, is presented. The search uses \SI{36.1}{\per\femto\barn} \pp{} collision data at a centre-of-mass energy of \SI{13}{\TeV} collected by the ATLAS experiment at the Large Hadron Collider and is performed in final states with jets identified as containing charm hadrons. Assuming a 100\% branching ratio to $c\ninoone$, top and charm squarks with masses up to \SI{850}{\GeV} are excluded at 95\% confidence level for a massless lightest neutralino. For  $m_{\stopone,\scharmone}-\mn <$~\SI{100}{\GeV}, top and charm squark masses up to \SI{500}{\GeV} are excluded.
}

\author{The ATLAS Collaboration}

\AtlasRefCode{SUSY-2016-26}

\PreprintIdNumber{CERN-EP-2018-034}

\AtlasJournal{JHEP}
\AtlasJournalRef{JHEP 09 (2018) 050}
\AtlasDOI{10.1007/JHEP09(2018)050}

%% file: latex/sections/Introduction.tex
Supersymmetry (SUSY)~\cite{Golfand:1971iw,Volkov:1973ix,Wess:1974tw,Wess:1974jb,Ferrara:1974pu,Salam:1974ig} provides an extension of the Standard Model (SM) which solves the hierarchy problem~\cite{Sakai:1981gr,Dimopoulos:1981yj,Ibanez:1981yh,Dimopoulos:1981zb} by introducing partners of the known bosons and fermions. In $R$-parity-conserving supersymmetric models~\cite{Farrar:1978xj}, supersymmetric partners are produced in pairs, and the lightest supersymmetric particle (LSP) is stable. The LSP could provide a viable candidate for dark matter~\cite{Goldberg:1983nd,Ellis:1983ew}. The scalar partners of right-handed and left-handed quarks, the $\tilde{q}_{\mathrm R}$ and  $\tilde{q}_{\mathrm L}$ squarks, can mix to form two mass eigenstates, $\tilde{q}_1$ and  $\tilde{q}_2$, with $\tilde{q}_1$ defined to be the lighter state. The charginos and neutralinos are mixtures of the binos, winos and Higgsinos that are the supersymmetric partners of the \Uone{} and \SUtwo{} gauge bosons and the Higgs bosons, respectively. Their mass eigenstates are referred to as $\tilde{\chi}^\pm_i$ $(i=1,2)$ and $\tilde{\chi}^0_j$ $(j=1,2,3,4)$ in order of increasing masses. In the present work, the LSP is assumed to be the lightest neutralino $\tilde{\chi}^0_1$.

\begin{figure}[h]
   \begin{center}
   \includegraphics[width=0.45\textwidth]{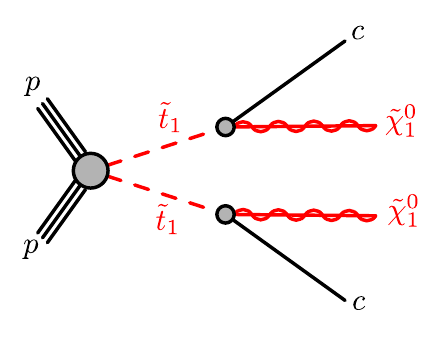}
   \includegraphics[width=0.45\textwidth]{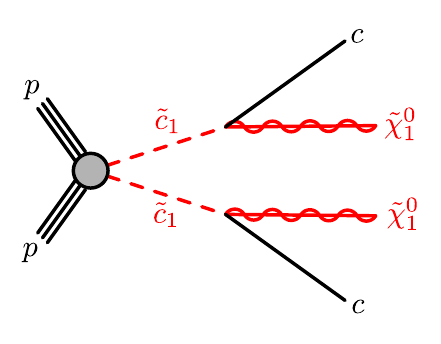}
   \end{center}
   \caption{
   Production diagrams for the pair production of top and charm squarks with subsequent decay into charm quarks and two LSPs. 
   The differentiation of particles and anti-particles is omitted in the diagrams but it is implicitly assumed. The direct squark production processes produce a top and charm squark--antisquark pair, respectively. 
   }
   \label{fig:feynmandiags}
\end{figure}

In supersymmetric models, naturalness~\cite{Barbieri:1987fn,deCarlos:1993yy} arguments suggest that the mass of the lightest top squark, $\tilde{t}_1$, should have mass of the same order of the top quark~\cite{Inoue:1982pi,Ellis:1983ed}.  The present analysis considers top-squark pair production with subsequent decays induced by flavour-violating effects into a charm quark and a LSP, $\stopone\to c\ninoone$, as indicated in the left panel of Figure~\ref{fig:feynmandiags}. This results in final-state signatures characterised by the presence of two charm quarks and missing transverse momentum from the LSPs which escape the detector. The magnitude of flavour-violating effects inducing $\stopone\to c\ninoone$ decays is model dependent, and in certain cases the branching ratio of the $\stopone\to c\ninoone$ decay can exceed the branching ratios of the three- and four-body decays ($\stopone\to Wb\ninoone$ and $\stopone\to bf\bar{f}'\ninoone$, respectively, where $f$ and $f'$ are two fermions from the virtual $W$ boson decay). This can occur due to significant mixing between the top squark and the charm squark, which may be induced by renormalisation group equations even if minimal flavour violation is assumed at a certain scale ~\cite{Grober:2015fia}. 

In generic supersymmetric models, squarks are not required to be mass degenerate, and in particular, charm squarks could be considerably lighter than other squark flavours~\cite{Mahbubani:2012qq}. This motivates searches for pair production of charm squarks, each of which decays as $\scharmone\to c\ninoone$, as illustrated in the right panel of Figure~\ref{fig:feynmandiags}. The experimental signature and the cross-section of such models are identical to those of top-squark production with the top squark decaying as $\stopone\to c\ninoone$ and, therefore, the two processes are treated on the same footing in this analysis.
For both processes, a 100\% branching ratio to $c\ninoone$ is assumed. 

This paper presents a search for pair production of top and charm squarks decaying into a charm quark and the lightest neutralino, using \intlumi~of proton--proton ($pp$) collisions data at a centre-of-mass energy of $\sqrt{s}=13$~\TeV\ collected by the ATLAS experiment at the Large Hadron Collider (LHC) in 2015 and 2016. A dedicated algorithm is used to identify jets containing charm hadrons (\cjets{}).
The search improves on previous results~\cite{Aad:2014nra,Aad:2015gna} from the ATLAS Collaboration using 20.3 ${\mathrm fb}^{-1}$ of 8 TeV data. A related search was recently reported by the CMS Collaboration~\cite{Sirunyan:2017kiw}.

%% file: latex/sections/Detector.tex
%-------------------------------------------------------------------------------
\newcommand{\AtlasCoordFootnote}{%
ATLAS uses a right-handed coordinate system with its origin at the nominal interaction point (IP)
in the centre of the detector and the $z$-axis along the beam pipe.
The $x$-axis points from the IP to the centre of the LHC ring,
and the $y$-axis points upwards.
Cylindrical coordinates $(r,\phi)$ are used in the transverse plane, 
$\phi$ being the azimuthal angle around the $z$-axis.
The pseudorapidity is defined in terms of the polar angle $\theta$ as $\eta = -\ln \tan(\theta/2)$. Rapidity is defined as $y = 0.5\ln[(E+p_z)/(E-p_z)]$, where $E$ denotes the energy and $p_z$ is the component of the momentum along the 
beam direction. Angular distance is measured in units of $\Delta R \equiv \sqrt{(\Delta y)^{2} + (\Delta\phi)^{2}}$.}
%-------------------------------------------------------------------------------

The ATLAS experiment~\cite{PERF-2007-01} at the LHC is a multi-purpose particle detector
with a forward-backward symmetric cylindrical geometry and a near $4\pi$ coverage in 
solid angle.\footnote{\AtlasCoordFootnote}
It consists of an inner tracking detector surrounded by a thin superconducting solenoid
providing a \SI{2}{\tesla} axial magnetic field, electromagnetic and hadron calorimeters, and a muon spectrometer.
The inner tracking detector covers the pseudorapidity range $|\eta| < 2.5$.
It consists of silicon pixel, silicon microstrip, and transition-radiation tracking detectors.
The newly installed innermost layer of pixel sensors \cite{ATLAS-TDR-19} was operational for the first time during the 2015 data-taking, leading to an improvement of the flavour-tagging performance \cite{ATL-PHYS-PUB-2016-012}.
Lead/liquid-argon (LAr) sampling calorimeters provide electromagnetic (EM) energy measurements
with high granularity.
A hadron (steel/scintillator-tile) calorimeter covers the central pseudorapidity range ($|\eta| < 1.7$).
The endcap and forward regions are instrumented with LAr calorimeters
for both the EM and hadronic energy measurements up to $|\eta| = 4.9$.
The muon spectrometer surrounds the calorimeters and is based on
three large air-core toroidal superconducting magnets with eight coils each.
The field integral of the toroids ranges between \num{2.0} and 6.0 Tm
across most of the detector. 
The muon spectrometer includes a system of precision tracking chambers and fast detectors for triggering.
A two-level trigger system~\cite{Aaboud:2016leb} is used to select events.
The first-level trigger is implemented in hardware and uses a subset of the detector information to reduce the accepted rate to at most \SI{100}{\kilo\hertz}.
This is followed by a software-based trigger level that reduces the accepted event rate to about \SI{1}{\kilo\hertz}.

%% file: latex/sections/DataAndMC.tex
The data used in this analysis were collected by the ATLAS detector in \pp{} collisions at the LHC with $\sqrt{s} = $~\SI{13}{\TeV} and \SI{25}{\ns} proton bunch crossing interval over the 2015 and 2016 data-taking periods.
The full dataset corresponds to an integrated luminosity of \SI{36.1}{\per\femto\barn} after the application of beam, detector, and data-quality requirements. 
The uncertainty in the combined 2015+2016 integrated luminosity is 2.1\%. It is derived, following a methodology similar to that detailed in Ref.~\cite{DAPR-2013-01}, from a calibration of the luminosity scale using $x$$-$$y$ beam-separation scans performed in August 2015 and May 2016.
On average, there are about 14 or 25 inelastic \pp{} collisions (``\pileup{}'') in each bunch crossing in the 2015 and 2016 datasets, respectively.  

In all search regions, events are required to pass a missing transverse momentum (\met{}) trigger with an online threshold of \SI{70}{\GeV} or \SI{110}{\GeV} for the 2015 and 2016 data periods, respectively. This trigger is fully efficient for events passing the preselection defined in Section \ref{sec:preselection}. Events in control regions are selected using triggers requiring the presence of one electron or muon. 

Monte Carlo (MC) simulated event samples are used to model the SUSY signal and to help in the estimation of SM background processes. 
The SUSY signal samples used in this analysis were generated using \MGMCatNLO{} v2.2.3 \cite{Alwall:2014hca} and interfaced to \PYTHIA{} 8.186 \cite{Sjostrand:2007gs} with the A14 set of tuned parameters \cite{ATL-PHYS-PUB-2012-003} for the simulation of parton showering, hadronisation, and the description of the underlying event. 
The matrix element (ME) calculation was performed at tree level includes the emission of up to two additional partons. The ME to parton shower (PS) matching was done using the \textsc{CKKW-L} prescription \cite{Lonnblad:2012ix}, with a matching scale set to $m_{\stopone,\scharmone}/4$. The mass of the top quark was fixed at \SI{172.5}{\GeV}. 
The \textsc{NNPDF2.3LO} \cite{Ball:2012cx} parton distribution function (PDF) set was used. 
Signal cross-sections were calculated at next-to-leading-order accuracy (NLO) in the strong coupling constant, adding the resummation of soft-gluon emission at next-to-leading-logarithm accuracy (NLO+NLL) \cite{Beenakker:1997ut,Beenakker:2010nq,Beenakker:2011fu}.  The nominal cross-section and its uncertainty are based on predictions using different PDF sets and factorisation and renormalisation scales, as described in Ref.~\cite{Borschensky:2014cia}. The same cross-section values were used for the pair production of top and charm squarks.

SM background samples were simulated using different MC event generator programs depending on the process. 
Events containing \W{} or \Z{} bosons with associated jets, and diboson events ($WW, WZ$ and $ZZ$) were simulated using the \SHERPA{} 2.2.1 \cite{Gleisberg:2008ta} event generator. The ME calculation is performed using \Comix{} \cite{Gleisberg:2008fv} and \OpenLoops{} \cite{Cascioli:2011va} for up to two partons at NLO and four partons at LO in \W{}/\Z{} + jets events and up to one parton at NLO and three partons at LO in diboson events, and merged with the \SHERPA{} parton shower \cite{Schumann:2007mg} according to the ME+PS@NLO prescription \cite{Hoeche:2012yf}. The \textsc{NNPDF3.0NNLO} PDF set \cite{Ball:2014uwa} was used in conjunction with dedicated parton shower tuning developed by the \SHERPA{} authors. The \W{}/\Z{} + jets events were normalised with NNLO QCD cross-sections~\cite{Gavin:2010az}. 
For the generation of \ttbar{} events, \POWHEGBOX v2 \cite{Alioli:2010xd} was used with the \textsc{CT10} PDF \cite{Lai:2010vv} set in the matrix element calculations. Electroweak $t$-channel, $s$-channel and $Wt$-channel single-top-quark events were generated with \POWHEGBOX v1. For $s$-channel and $Wt$-channel production, the CT10 PDF set was used, while for the $t$-channel production the four-flavour scheme was used for the NLO matrix element calculations together with the fixed four-flavour PDF set CT10f4. For \ttbar{} and single-top-quark processes, top-quark spin correlations were preserved.  
The parton shower, hadronisation, and the underlying event were simulated using \PYTHIA{} 6.428 \cite{Sjostrand:2006za} with the \textsc{CTEQ6L1} PDF set and the corresponding Perugia 2012 set of tuned parameters~\cite{Skands:2010ak}. The top-quark mass was set to \SI{172.5}{\GeV}. At least one leptonically decaying \W{} boson was required in all generated \ttbar{} and single-top-quark events. Fully hadronic decays do not produce sufficient \met{} to significantly contribute to the background. 

The production of top-quark pairs in association with $W, Z$ or Higgs bosons was modelled by samples generated at NLO using \madgraph~v2.2.3 and showered with \PYTHIA{} 8.212. The $t\bar{t}WW$ process and the production of events with three or four top quarks were modelled at leading order (LO) using \madgraph~v2.2.3 interfaced to \PYTHIA{} 8.212. 

The \EvtGen{1.2.0} program \cite{EvtGen} was used by all generators except \SHERPA{} to describe the properties of $b$- and $c$-hadron decays. 

Additional samples from different event generators and settings are used to estimate systematic uncertainties in the backgrounds, as described in Section \ref{sec:uncert}.

All simulated events were overlaid with multiple \pp{} collisions simulated with the soft QCD processes of \PYTHIA{} 8.186 with the A2 tune settings and the \textsc{MSTW2008LO} PDF set \cite{Martin:2009iq}, and reweighted to match the \pileup{} distribution observed in the data. 
The MC samples were processed through the ATLAS full detector simulation \cite{SOFT-2010-01} based on \GEANT{4} \cite{Agostinelli:2002hh} or a fast simulation \cite{ATL-PHYS-PUB-2010-013} that uses a parameterisation for the calorimeter response and \GEANT{4} for the rest of the detector. 
All simulated events are reconstructed with the same algorithms as used for the data. 
In order to improve agreement between data and simulation, corrections are applied to the lepton trigger and identification efficiencies, momentum scale, energy resolution, and the efficiency of identifying jets containing charm hadrons~\cite{ATLAS-CONF-2016-024,PERF-2013-05,PERF-2015-10}.

%% file: latex/sections/EventReco.tex
Interaction vertices from the proton--proton collisions are reconstructed from at least two tracks with transverse momentum  \pt{}~$>$~\SI{400}{\MeV}, and are required to be consistent with the beam-spot envelope. The primary vertex is identified as the one with the largest sum of squares of the transverse momenta from associated tracks, $\sum p_{\mathrm{T,track}}^2$ \cite{ATL-PHYS-PUB-2015-026}.

Reconstructed physics objects are labelled as \textit{baseline} or \textit{signal}, with the latter satisfying tighter identification and quality criteria. Baseline objects are used for the removal of overlaps between different reconstructed objects (``overlap removal'') and for lepton-veto purposes. 

Electron baseline candidates are reconstructed from energy clusters in the electromagnetic calorimeter matched to inner detector tracks.  A multivariate calibration is applied to correct the electron energy scale \cite{PERF-2013-05}. Electron candidates must have $|\eta|<$~2.47 and $\pt>$~\SI{7}{\GeV}, and meet a set of ``loose'' quality criteria \cite{ATLAS-CONF-2016-024}. 

Muon baseline candidates are reconstructed from tracks in the muon spectrometer and the inner detector. Candidates for which the inner detector tracks are matched to an energy deposit in the calorimeter compatible with a minimum-ionizing particle are also considered. Muon candidates must have $|\eta|<$~2.7 and $\pt>$~\SI{7}{\GeV}, and meet a set of ``loose'' quality criteria \cite{PERF-2015-10}. 

For signal lepton candidates used in the control region definitions, additional requirements are applied. Signal electrons are required to meet a set of ``medium'' quality criteria \cite{ATLAS-CONF-2016-024}, a ``loose'' isolation requirement \cite{ATLAS-CONF-2016-024}, and to have $\pt>$ \SI{27}{\GeV}. Signal muons are required to pass a ``loose'' isolation requirement \cite{PERF-2015-10} and to have $\pt>$ \SI{27}{\GeV}.

Jets are reconstructed from three-dimensional topological clusters \cite{PERF-2014-07} of calorimeter cells using the \antikt algorithm \cite{Cacciari:2008gp} with a radius parameter of $R=0.4$ and a four-momentum recombination scheme. The jet energy is then corrected to the particle level by the application of a jet energy scale calibration derived from \SI{13}{TeV} data and simulation \cite{PERF-2016-04}. To reduce \pileup{} effects, jets with $|\eta|<2.4$ are tested for compatibility with the primary vertex and required to pass a set of ``medium'' requirements \cite{PERF-2014-03}. 
Baseline jets are required to have $|\eta|<$ 2.8 and $\pt>$~\SI{20}{\GeV}. These jets are used in the overlap removal procedure. Signal jets are required to also have $|\eta|<$~2.5 and $\pt>$~\SI{30}{\GeV}. 

Identifying \cjets{} is of crucial importance to this analysis. Two multivariate discriminants, MV2c100 and MV2cl100, are used to distinguish between \cjets and jets containing bottom hadrons (\bjets), and between \cjets and 
light-flavour jets (jets not containing a $b$- or a $c$-hadron), respectively. The discriminants are based on the MV2 algorithm presented in 
Refs.~\cite{ATL-PHYS-PUB-2015-039,ATL-PHYS-PUB-2016-012}. MV2c100 is trained with \bjets as signal 
and with background consisting exclusively of \cjets. MV2cl100 is trained with \cjets as signal and with 
background consisting exclusively of light-flavour jets.
A ``tight'' working point was adopted in the analysis, resulting in a charm-tagging efficiency of about 18\%, a $b$-jet rejection factor of 20, a light-flavour jet rejection factor of 200, and a hadronic $\tau$ jet rejection factor of 6, as evaluated in a sample of simulated $t\bar{t}$ events.
The \cjet tagging rate is measured using a data sample rich in $t\bar{t}$ events, identifying \cjets from $W$ boson decays. Correction factors dependent on the jet \pt are derived to match the tagging rate in the simulation to that measured in data. They are found to range from 0.83 to 0.88 in \PYTHIA{} 8 MC samples. 
Similarly, correction factors for the mis-tag rate for $b$-jets are derived using dileptonic \ttbar{} data events. They are found to range from 0.81 to 0.97  in \PYTHIA{} 8 MC samples. 
Correction factors for the mis-tag rate for light-flavour jets are instead evaluated by comparing the MC simulation to an ``adjusted'' simulation where the track-based variables used in the flavour-tagging evaluation, e.g. the transverse impact parameter relative to the primary vertex, are scaled to match the data. 
The size of this correction is found to range between 1.3 and 1.9 across the considered jet \pt and $\eta$ ranges in \PYTHIA{} 8 MC samples.  

To avoid object double counting, an overlap removal procedure is applied to resolve ambiguities among electrons, muons, and jets in the final state. 
Any muon candidate that is reconstructed with inner detector and calorimeter information only and which shares an inner detector track with an electron candidate is removed. Any electron candidate sharing an inner detector track with a remaining muon candidate is also removed. Jets within $\DR = 0.2$ of an electron are discarded. Jets with less than three tracks within $\DR = 0.2$ of a muon candidate are discarded. Any remaining electron candidate within $\DR = 0.4$ of a jet is discarded. Any remaining muon candidate within $\DR = 0.04 + 10/\pt^{\mu}$(\GeV) of a jet is discarded. 

The missing transverse momentum \metv{}, whose magnitude is denoted by \met{}, is defined as the negative vector sum of the transverse momenta of electrons, jets and muons, to which is added a ``soft term'' that recovers the contributions from other low-\pt particles \cite{Aad:2016nrq, Aaboud:2018tkc}. The soft term is constructed from all tracks that are not associated with any of the preceding objects, and that are associated with the primary vertex. 

%% file: latex/sections/EventSelection.tex
All events used in this analysis are required to satisfy basic quality criteria in order to guarantee the good functionality of the detector during the data-taking and to increase the purity and quality of the data sample by rejecting non-collision events such as cosmic-ray and beam-halo events. To ensure that the event contains a $pp$ interaction with large transverse momentum transfer, at least one vertex with a position consistent with the beam-spot position is required. 

All events in signal regions are required to pass a preselection, presented in Section \ref{sec:preselection}, that targets the \cjets{} + \met{} final state. Variables designed to reject SM background events are also defined in Section \ref{sec:preselection}. Tighter selections are then applied to maximise the sensitivity to the signal models of interest, resulting in five signal regions, described in Section \ref{sec:SRs}. 

\subsection{Preselection}
\label{sec:preselection}

Stop and scharm pair production followed by \StopInCharm{} and \ScharmInCharm{} decays have the same final state characterised by \met{} and at least two \cjets{}. 
No prompt leptons are expected, and therefore events with at least one baseline electron or muon candidate, as defined in Section \ref{sec:eventReco}, are vetoed in order to reduce contamination from SM background processes 
with at most a signal rejection of only a few percent. 
In order to ensure full efficiency for the \met{} trigger used in the data selection, a requirement on \met{}~$>$~\SI{250}{\GeV} is applied. 
At least two jets are required, and the jet leading in \pt (leading jet) is required to satisfy $\pt^{j_1}>$~\SI{250}{\GeV}. In addition, at least one jet is required to be charm-tagged. A requirement on the minimum angle between the \metv{} and each signal jet, $\DphiJetMet{}>$ 0.4, is imposed in order to reduce the multijet background arising from mismeasured \met{}.

\subsection{Signal Regions}
\label{sec:SRs}
The signal characteristics depend strongly on the mass difference between the top or charm squark and the lightest neutralino, $\Dm=m_{\stopone,\scharmone}-\mn$. 
A large mass difference ($\Dm{}~>$~\SI{300}{\GeV}) leads to high-\pt{} jets and high \met{}. For smaller mass differences, the \met{} is substantially reduced. To target such signal scenarios, an additional initial-state radiation (ISR) jet is required to boost the system, leading to larger values of \met{}. 
Specific signal region selections (SRs) are optimised to target different signal topologies. Five signal regions (SR1--SR5) are defined in total, targeting scenarios with increasingly large mass splittings $\Delta m$. All signal region selections require a missing transverse momentum of at least \SI{500}{\GeV}.

Requiring the presence of one or more charm-tagged jets results in an enhancement of events with hadronically decaying $\tau$-leptons, as they have a high probability of being mis-tagged. The minimum transverse mass \mtc{} of charm-tagged jets and \met{} is defined as

\begin{equation*}
  \mtc = \min_{\text{$c$-jets}} \sqrt{2\cdot \met p_\mathrm{T}^{c} \cdot (1-\cos{\Delta\phi( \metv{} , \textit{\textbf{p}}_\mathrm{T}^{c})}) },
  \label{eq:mtc}
\end{equation*}
where $\Delta\phi( \metv{} , \textit{\textbf{p}}_\mathrm{T}^{c})$ is the azimuthal angle between the charm-tagged jet and the missing transverse momentum. The $\mtc$ variable shows a large rejection power against events with hadronically decaying $\tau$-leptons. The requirement of $\mtc>$\SI{120}{\GeV} in all SRs reduces this contribution to less than 5\% of the total background events. 
The \mtc{} variable also provides discrimination against other backgrounds, such as $Z$+jet production with invisibly decaying $Z$ bosons. 

An overview of the selection requirements in all signal regions is given in Table \ref{table:SRDefinitions}. 

Signal region {\textbf{SR1}} targets compressed mass scenarios with \Dm{} smaller than $\sim$\SI{50}{\GeV}. In such scenarios, the \cjets{} are expected to have \pt{} below the \SI{30}{\GeV} threshold for signal jets, and may not always be reconstructed. 
For even smaller values of \Dm{}, neither \cjet{} is expected to be reconstructed. 
Another analysis, the ``monojet'' analysis described in Ref.~\cite{Aaboud:2017phn}, is sensitive to this type of signal. Signal region {\textbf{SR2}} targets compressed mass scenarios with \Dm{} larger than \SI{50}{\GeV} but smaller than $\sim$\SI{100}{\GeV}. Both \cjets{} are expected to be reconstructed and at least three jets are required.  Signal regions {\textbf{SR3}} and {\textbf{SR4}} target scenarios with intermediate values of $\Dm$, 100--\SI{450}{\GeV}. Requirements on the charm-tagged jet transverse momentum and on \mtc{} are tightened, and more stringent thresholds on the second and third jets' transverse momenta are applied. An upper bound is imposed on the value of \mtc{} in signal regions {\textbf{SR1}--\textbf{SR3}} as this is found to increase the sensitivity of the analysis. In signal regions {\textbf{SR1}--\textbf{SR4}}, the leading jet is primarily expected to arise from initial-state radiation, and is therefore required to fail the charm-tagging requirement. Finally, signal region {\textbf{SR5}} targets scenarios with values of \Dm\ greater than around $450$~\GeV, such that the leading two jets are usually the \cjets{} from the decay of the \stopone or \scharmone. Therefore, the leading jet is no longer required to fail the charm-tagging requirement. In addition, requirements on the jet transverse momenta and \mtc{} are further tightened relative to {\textbf{SR1}--\textbf{SR4}}. 

According to the selected benchmark model, about 20 signal events pass the selection criteria for the corresponding SR of interest. A full overview is reported in Table \ref{tab:SRyields}. 

\begin{table}[h]
  \begin{centering}
    \begin{tabular}{l|c|c|c|c|c} 
      & SR1 & SR2 & SR3 & SR4 & SR5 \\
      \hline
      \hline
      	Trigger & \multicolumn{5}{c}{\met{} triggers} \\
      	Leptons & \multicolumn{5}{c}{0 $e$ \texttt{AND} 0 $\mu$ } \\
      \met [GeV]           &  \multicolumn{5}{c}{$>500$} \\
      \DphiJetMet{} [rad]           &   \multicolumn{5}{c}{$> 0.4$}\\
      \nctag{}            &   \multicolumn{5}{c}{$\ge1$}\\
      \hline
      \njets{}                 & $\geq 2$     & $\geq 3$ & $\geq 3$         & $\geq3$   & $\geq 3$ \\
      Leading jet $c$-tag veto & yes          & yes      & yes              & yes       &     no     \\
      \jetOnePt [GeV]      & $>250$ & $>250$ & $>250$ & $>250$ & $>300$   \\
      \jetTwoPt [GeV]      &     $-$     &    $-$      & $>100$           & $>140$    & $>200$   \\
      \jetThreePt [GeV]    &    $-$          &     $-$     & $>80$            & $>120$    &  $>150$        \\
      \cTagOnePt [GeV]     & $<100$       & $>60$    & $>80$            & $>100$    & $>150$   \\
      \mtc\ [GeV]           & $\in(120, 250)$       & $\in(120, 250)$   & $\in(175, 400)$ & $>200$   & $>400$   \\
    \end{tabular}
  \par\end{centering}{\tiny \par}
  \caption{\label{table:SRDefinitions}
  Overview of the signal region selection criteria. 
  \njets{} and \nctag{} indicate the total number of jets and \cjets{}, respectively; 
  \jetOnePt{}, \jetTwoPt{} and \jetThreePt{} indicate the transverse momentum of the leading, sub-leading and sub-sub-leading jet; and 
  \cTagOnePt{} is the transverse momentum of the \cjet{} with the highest \pt{}. 
  }
\end{table}

%% file: latex/sections/BackgroundEstimation.tex
The main background contribution in this analysis arises from \Z{}+jets production with the \Z{} boson decaying into neutrinos. 
Across all the SRs this process contributes up to 50--60\% of the total background expectation. 
Other significant background contributions arise from \W{}+jets production in association with a heavy-flavour jet where the \W{} boson decays into a $\tau$-lepton and a neutrino ($\sim$10\% of the total background expectations in the SRs), followed by diboson and \ttbar{} production ($\sim$10\% and 5--10\% of the total SR expectations, respectively).
Minor backgrounds including \Ztautau{}, single-top and \ttbar{} production in association with additional vector bosons, which account for less than 5\% of the total background, are grouped together and hereafter referred to as the ``Other'' contribution. 
The contribution of the multijet background is evaluated using dedicated data-driven techniques, but is found to be negligible in all regions and therefore not further considered. Contributions from diboson production and other small backgrounds are determined from simulation only.
   
The \Z{}+jets, \W{}+jets and top-quark background yields in the SRs are estimated using simulated samples normalised to data in dedicated control regions (CRs). Three CRs are constructed separately for each SR. They are designed to enhance the contribution of a particular background component, as described in Section \ref{sec:CR}, and are required to be disjoint from the SRs while still retaining similar kinematic and flavour-composition properties.

\subsection{Control region definitions}
\label{sec:CR}

To ensure that the control regions are statistically independent of the signal regions, the lepton veto applied in the preselection (Section \ref{sec:preselection}) is modified such that each control region requires at least one signal electron or muon. 
In order to ensure a similar background composition between signal and control regions, identical requirements are placed on the jet multiplicity and flavour tagging in control regions and corresponding signal regions.
Requirements on \met{} and jet momenta are lowered in the control regions relative to the signal regions, to ensure that the control region samples have enough events. In addition to these common criteria, specific selections to target \Z{}+jets, \W{}+jets and top-quark production backgrounds are imposed as described in Sections \ref{sec:ZCR} and \ref{sec:WTCR}. The potential contamination from signal in each CR is found to be negligible for all signal hypotheses that have not yet been excluded.

\subsubsection{\Z{} control regions}
\label{sec:ZCR}

The \Z{}+jets events that pass the SR requirements are mostly from $Z(\to\nu\nu)$ production in association with a charm quark, followed by $Z(\to\nu\nu)$ production in association with light-flavour jets, some of which have been mis-tagged as $c$-jets.
Control regions enriched in $Z$+jets events are defined by requiring two same-flavour (SF), opposite-sign (OS) leptons (electrons or muons) with an invariant mass $\mll$ close to the \Z{} boson mass (\SI{76}{\GeV}~$<\mll<$~\SI{106}{\GeV}). 
The lepton momenta are then vectorially added to the \metv{} to mimic the expected missing transverse momentum spectrum of $Z(\to\nu\nu)$+jets events. 
Other quantities depending on \met{} are also recomputed accordingly. 
In the following, all variables corrected in this manner are indicated with the superscript ``corr''. A missing transverse momentum requirement \met{}~$<$~\SI{75}{\GeV} is also applied in the \Z{} control regions to reduce the contribution of the \ttbar{} background. 
The full set of criteria is summarised in Table \ref{tab:ZCR}. MC studies show that the jet flavour composition in all the \Z{} control regions is in excellent agreement with the composition in the signal regions.

\begin{table}[th]
  \begin{centering}
    \renewcommand{\arraystretch}{1.1}
    \begin{tabular}{l|c|c|c|c|c} 

      & \Z{} CR1 & \Z{} CR2 & \Z{} CR3 & \Z{} CR4 & \Z{} CR5 \\

      \hline
      \hline
      
Trigger       & \multicolumn{5}{c}{\met \texttt{OR} single lepton} \\
Leptons     & \multicolumn{5}{c}{2 SF OS $e$ \texttt{OR} $\mu$} \\ 
\mll{}	[GeV]	 & \multicolumn{5}{c}{$\in(76, 106)$}\\      
      \DphiJetMetCorr{} [rad]            &   \multicolumn{5}{c}{$> 0.4$}\\
      \nctag{}            &   \multicolumn{5}{c}{$\ge1$}\\ 
      \met{} [GeV]           &  \multicolumn{5}{c}{$<$ 75} \\
      \metcorr{} [GeV]           &  \multicolumn{5}{c}{$>$ 250} \\
      \jetOnePt{} [GeV]      &     \multicolumn{5}{c}{$>$ 250}  \\

     \hline

      \njets{}                 & $\geq 2$     & $\geq 3$ & $\geq 3$         & $\geq3$   & $\geq 3$ \\
      Leading jet $c$-tag veto & yes          & yes      & yes              & yes       &     no     \\
      \jetTwoPt [GeV]      &     $-$      &  $-$     & $>100$           & $>100$    & $>100$   \\
      \jetThreePt [GeV]   &    $-$       &  $-$     & $>80$            & $>100$    &       $-$   \\
      \cTagOnePt [GeV]     & $<100$       & $>60$    & $>80$            & $>100$    & $>100$   \\
      \mtccorr{} [GeV]           & $\in(120,250)$       & $\in(120,250)$   &  $\in(175, 400)$ & $>200$    & $>400$   \\
    \end{tabular}
  \par\end{centering}{\tiny \par}
  \caption{\label{tab:ZCR}
  Overview of the \Z{} control region selection criteria. 
  \njets{} and \nctag{} indicate the total number of jets and \cjets{}, respectively;
  \jetOnePt{}, \jetTwoPt{} and \jetThreePt{} indicate the transverse momentum of the leading, sub-leading and sub-sub-leading jet; and  
  \cTagOnePt{} is the transverse momentum of the \cjet{} with the highest \pt{}. 
 Variables with the superscript ``corr'' refer to variables corrected as described in the text.
  }
\end{table}

\subsubsection{\W{} and top-quark control regions}
\label{sec:WTCR}

The \W{}+jets and top-quark background events passing the SR requirements are mostly events with hadronically decaying $\tau$-leptons, where the charm-tagged jet is an ISR jet containing a charm hadron, or a mistagged light-flavour jet. Most of the events with a hadronically decaying $\tau$-lepton mis-tagged as a \cjet{} are instead rejected by the \mtc{} requirement applied in the SRs. 
Samples enriched in both the \W{}+jets and top-quark processes are obtained by selecting events with exactly one signal lepton (electron or muon) and with \W{} boson transverse mass $\mt = \sqrt{2\cdot \met \ptlepsc \cdot (1-\cos{\Delta\phi( \metv{} , \ptlepvec )}) }>$~\SI{60}{\GeV}.

From simulation it is determined that in the SRs on average 40\% of the $\tau$-lepton's momentum is carried by visible decay products, while the remaining 60\% is carried by neutrinos.
Therefore, in the CRs, to mimic the $\tau$-lepton's decay behaviour as closely as possible, single-lepton events are selected and the lepton is replaced with a jet whose \pt{} is 40\% of the lepton \pt{}, to mimic the visible part of the $\tau$-lepton decay. The remaining 60\% of the lepton momentum is added to the \met{} calculation, to mimic the neutrino contribution. 
Jets obtained using this procedure are further considered in the analysis only if they pass the signal jet $\pt$ and $\eta$ requirements described in Section \ref{sec:reco}. 
All variables computed from either jets or \met{} are recomputed accordingly and indicated with the superscript ``corr''.

As in the \Z{} CRs, the requirements on \met{} and jet momenta are relaxed relative to the corresponding signal regions to obtain enough events for the background estimation, while all other criteria are kept the same. 

In order to define separate CRs for \W{}+jets and the top-quark processes, two kinematic variables are introduced. 

\begin{description}
  \item{\mthad: }
    this variable is only defined in events with at least three jets. First, the jet pair with the invariant mass \mjj{} closest to the \W{} boson mass is selected. Then a third jet is selected such that the invariant mass \mjjj{} of all three jets is closest to the top-quark mass. In addition, at least one of these jets is required to be charm-tagged.
  \item{\mjjw: }
    the jet pair with \mjj{} closest to the \W{} boson mass is selected. Here, all pairs of jets where neither jet is charm-tagged are considered first, and the pair with \mjj{} closest to the \W{} boson mass is chosen. If no pairs of jets where neither jet is charm-tagged exist, then pairs containing at least one charm-tagged jet are considered and the pair with \mjj{} closest to the \W{} boson mass is chosen.
\end{description}

When computing \mthad and \mjjw, jets obtained from the replacement of leptons as described above are not considered. 
The two variables have a strong discrimination power between top-quark and \W{}+jets processes. The requirement of \SI{50}{\GeV}~$< \mthad{} <$~\SI{220}{\GeV} results in a 62--71\% pure top-quark sample, depending on the region. 
The \W{} CRs are defined by inverting the \mthad{} selection and requiring in addition $\mjjw{} >$~\SI{175}{\GeV}. This results in a 55--78\% purity for the \W{}+jets process, depending on the region. An overview of the full selection list for the \W{} CRs and the top-quark CRs (referred to as ``Top CRs'') are reported in Table \ref{tab:WTCR}. 
MC studies show that the jet flavour composition in all the \W{} control regions is in excellent agreement with that in the signal regions, except for \W{} CR4, where a difference up to 33\% is observed between SR and CR. An additional systematic uncertainty is derived to account for this difference as described in Section \ref{sec:uncert}.

\begin{table}[thb]
 \begin{centering}
  \renewcommand{\arraystretch}{1.3}
    \begin{tabular}{l|c|c|c|c|c} 
      & \W{}/Top CR1 & \W{}/Top CR2 & \W{}/Top CR3 & \W{}/Top CR4 & \W{}/Top CR5 \\
      \hline
      \hline
      Trigger          & \multicolumn{5}{c}{\met \texttt{OR} single lepton} \\
      Lepton          & \multicolumn{5}{c}{1 $e$ \texttt{OR} 1 $\mu$} \\ 
      \ptlepsc{} [GeV]    & \multicolumn{5}{c}{$ > 27$ }\\      
      \DphiJetMet{} [rad]   & \multicolumn{5}{c}{$\DphiJetMet{}>0.4$ \texttt{OR} $\DphiJetMetCorr{}>0.4$}\\
      \mt{} [GeV]    & \multicolumn{5}{c}{$ > 60$ }\\
      \metcorr [GeV]           &  \multicolumn{5}{c}{$>$ 250 } \\
      \jetOnePtCorr [GeV]      &  \multicolumn{5}{c}{$>$ 250 } \\
      \nctag{}            &   \multicolumn{5}{c}{$\ge 1$}\\
      \hline      
      \njetscorr{}                 & $\geq 2$     & $\geq 3$ & $\geq 3$   & $\geq3$   & $\geq 3$ \\
      Leading jet $c$-tag veto & yes          & yes      & yes              & yes       &     no     \\
      \jetTwoPtCorr [GeV]      &       $-$       &    $-$      & $>100$           & $>100$    & $>100$  \\
      \jetThreePtCorr [GeV]    &       $-$       &    $-$      & $>80$            & $>100$    &      $-$            \\
      \cTagOnePt [GeV]     & $<100$       & $>60$    & $>80$            & $>100$    & $>100$       \\
      \mtccorr\ [GeV]           & $\in(120,250)$       & $\in(120,250)$   & $\in(175, 400)$ & $>200$    & $>400$      \\
      \hline      
     \multirow{2}*{\mthad{} [GeV]}    &  \multicolumn{5}{c}{For \W{}: $< 50$  \texttt{OR} $> 220$ } \\
                                                        &  \multicolumn{5}{c}{For Top: $\in(50,220)$ }\\
      \hline      
      \multirow{2}*{\mjjw{} [GeV]}	& \multicolumn{5}{c}{For \W{}: $ > 175$ } \\
      & \multicolumn{5}{c}{For Top: no requirement} \\

    \end{tabular}
  \par\end{centering}{\tiny \par}
 \caption{\label{tab:WTCR} 
 Overview of the \W{} and Top control region selection criteria. 
  \njets{} and \nctag{} indicate the total number of jets and \cjets{}, respectively; 
  \jetOnePt{}, \jetTwoPt{} and \jetThreePt{} indicate the transverse momentum of the leading, sub-leading and sub-sub-leading jet; and  
  \cTagOnePt{} is the transverse momentum of the \cjet{} with the highest \pt{} and \ptlepsc{} is the lepton \pt{}.
 Variables with the superscript ``corr'' refer to variables corrected as described in the text.
 } 
\end{table}

\subsection{Background-only fit}
\label{sec:bkgFit}

For each signal region, the normalisations of the $W$+jets, $Z$+jets and top-quark backgrounds are determined through a binned likelihood fit performed simultaneously to the observed number of events in the associated control regions. The likelihood is constructed as a product of Poissonian probability density functions describing the observed data counts in the control regions and Gaussian constraint terms for the systematic uncertainties described in Section \ref{sec:uncert}. Poisson distributions are also included in the likelihood to describe the MC simulation's statistical uncertainties.
The fit is performed using the \text{HistFitter} package \cite{Baak:2014wma} with the assumption that no signal is present and it is referred to as the ``background-only'' fit. 
Correlations in background yield predictions across different regions due to common systematic uncertainties are accounted for in the fit. 

The background normalisation factors ($\mu$ factors) obtained in the fit are summarised in Table \ref{tab:mu}, including all the detector-related systematic uncertainties as described in Section \ref{sec:uncert}. 

\begin{table}[th]
\begin{centering}
\renewcommand{\arraystretch}{1.3}
\begin{tabular}{lcccccc}
Fit parameter & SR1 & SR2 & SR3 & SR4 & SR5 \\ 
\hline 
\hline
   $\mu_\W{}$                   & $1.13^{+0.38}_{-0.28}$ & $1.20^{+0.44}_{-0.39}$ & $1.70^{+0.39}_{-0.33}$ & $1.49^{+0.31}_{-0.26}$ & $1.28^{+0.30}_{-0.24}$ \\
   $\mu_\Z{}$                    & $1.36^{+0.17}_{-0.15}$ & $1.34^{+0.26}_{-0.22}$ & $1.22^{+0.27}_{-0.23}$ & $1.29^{+0.26}_{-0.22}$ & $1.30^{+0.28}_{-0.23}$ \\
   $\mu_{\mathrm{Top}}$  & $1.15^{+0.38}_{-0.28}$ & $1.28^{+0.55}_{-0.47}$ & $0.86^{+0.31}_{-0.35}$ & $0.79^{+0.35}_{-0.29}$ & $1.02^{+0.46}_{-0.32}$ \\
\end{tabular}
\caption{\label{tab:mu} Normalisation factors after the background-only fit. The central values with the detector-related systematic uncertainties, corresponding to one standard deviation, are reported.} \par\end{centering}{\tiny\par}
\end{table}

\begin{figure}[h]\centering
   \subfloat[\label{fig:ZCR1}]{
     \includegraphics[width=0.49\textwidth]{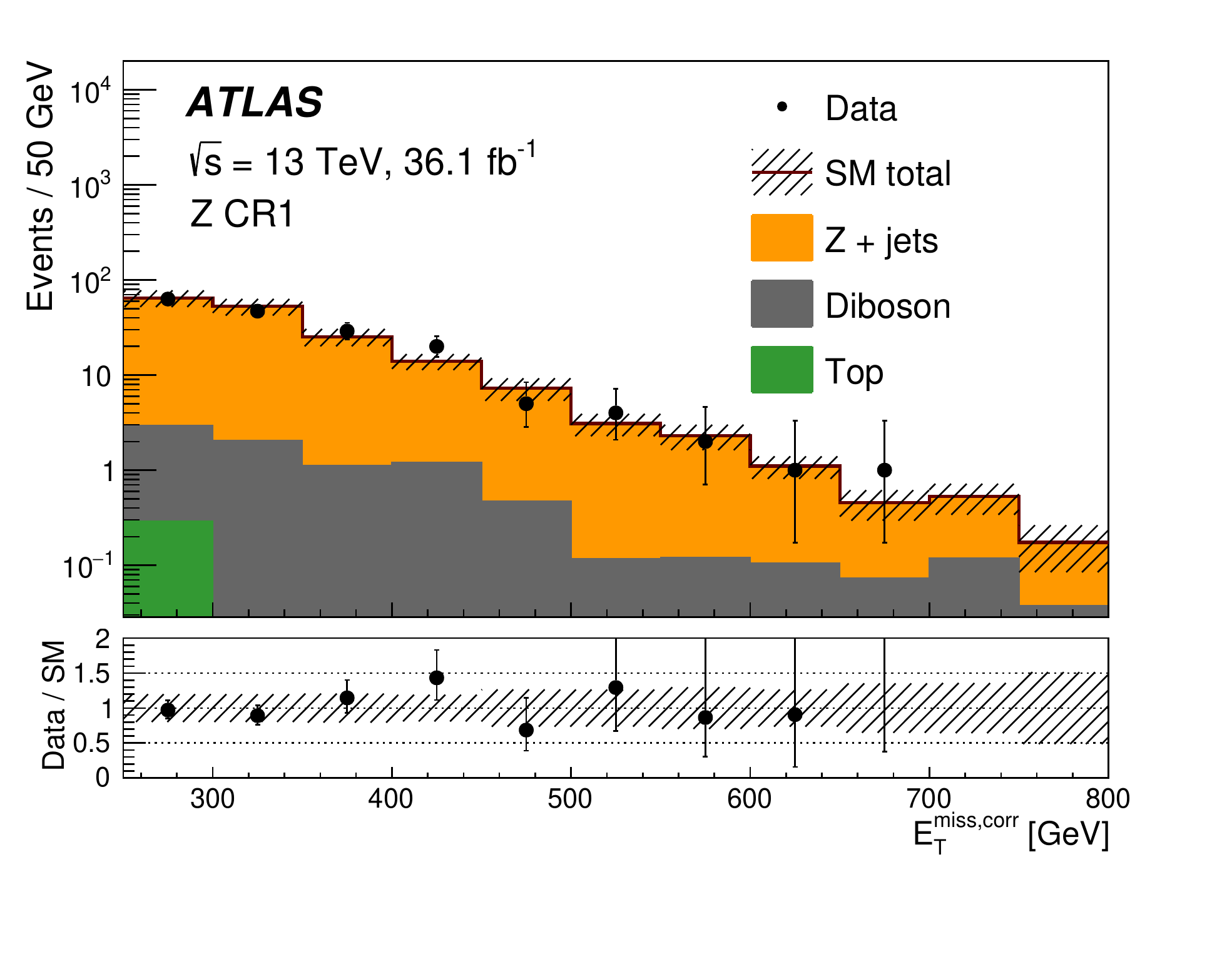}
   }
   \subfloat[\label{fig:ZCR5}]{
    \includegraphics[width=0.49\textwidth]{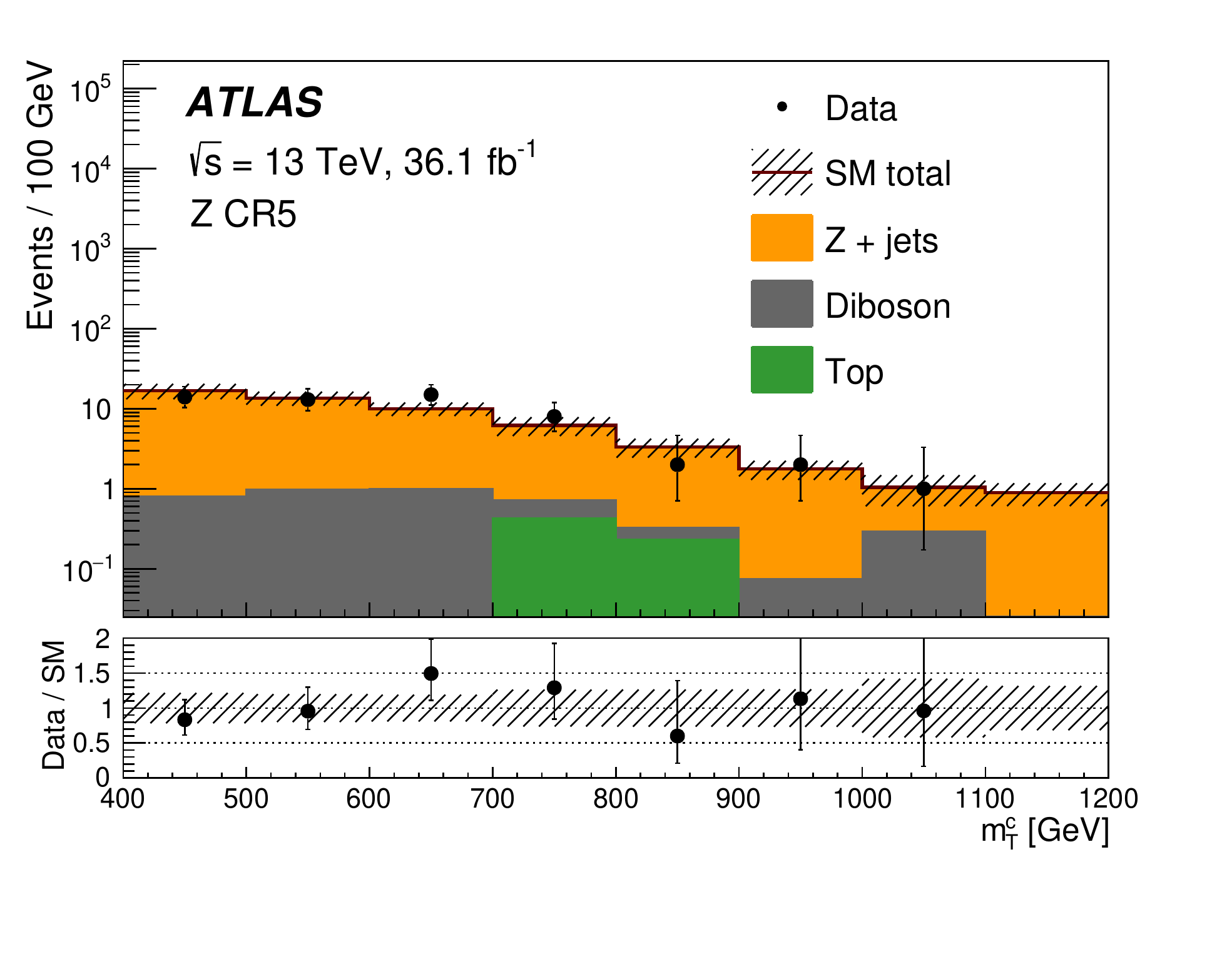}
   }\\
   \subfloat[\label{fig:WCR2}]{
    \includegraphics[width=0.49\textwidth]{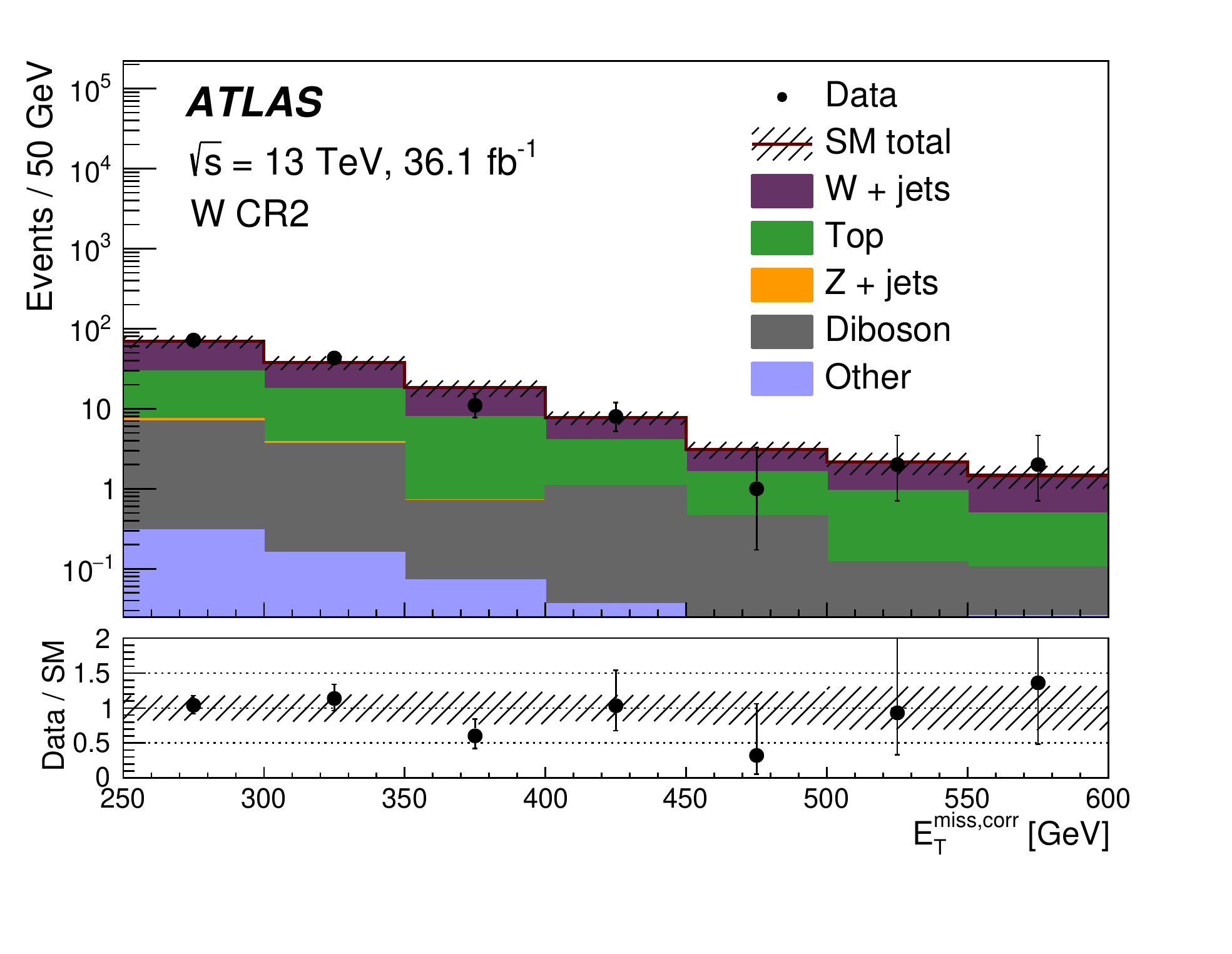}
   }
   \subfloat[\label{fig:WCR4}]{
    \includegraphics[width=0.49\textwidth]{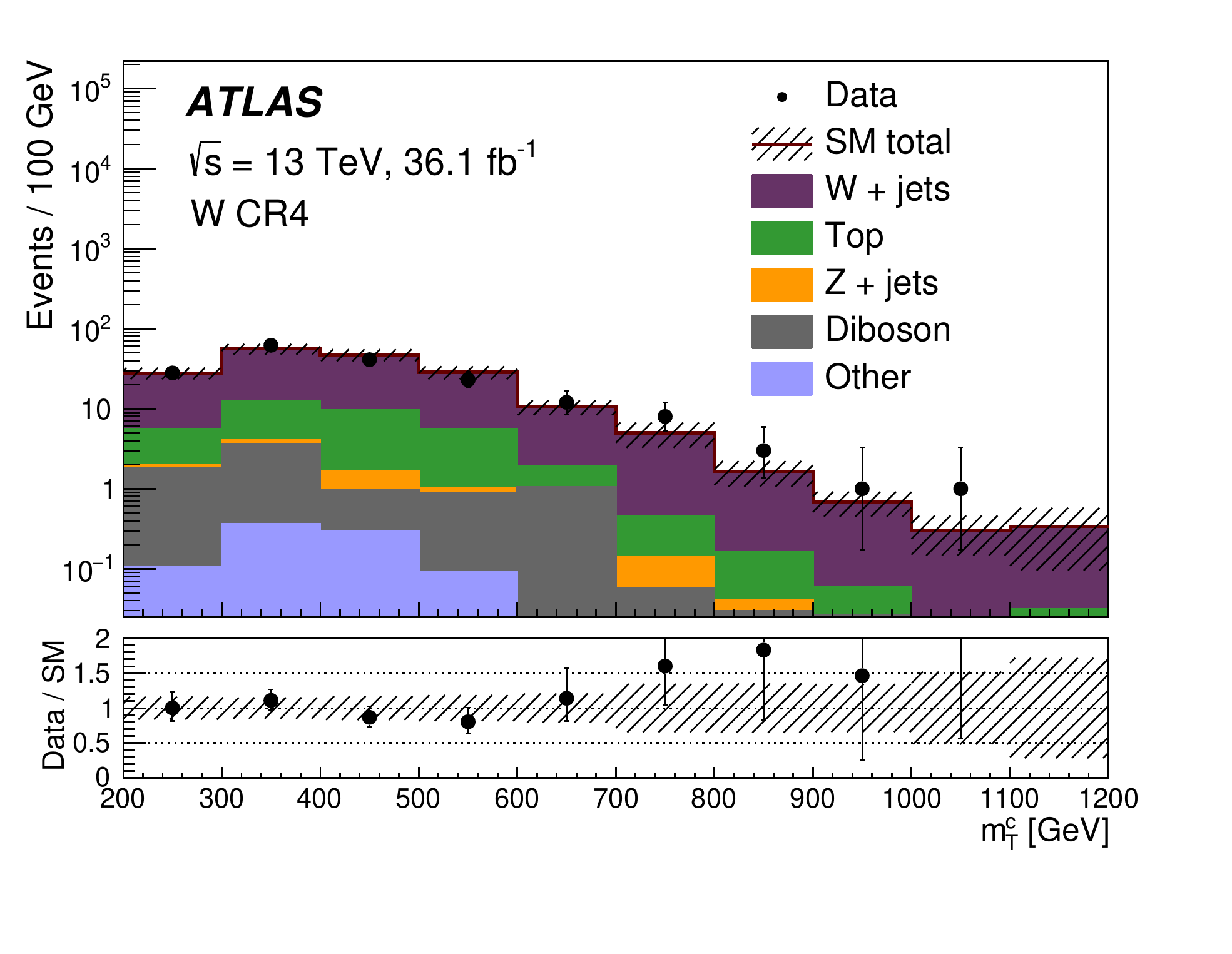}
   }\\
   \subfloat[\label{fig:TopCR3}]{
    \includegraphics[width=0.49\textwidth]{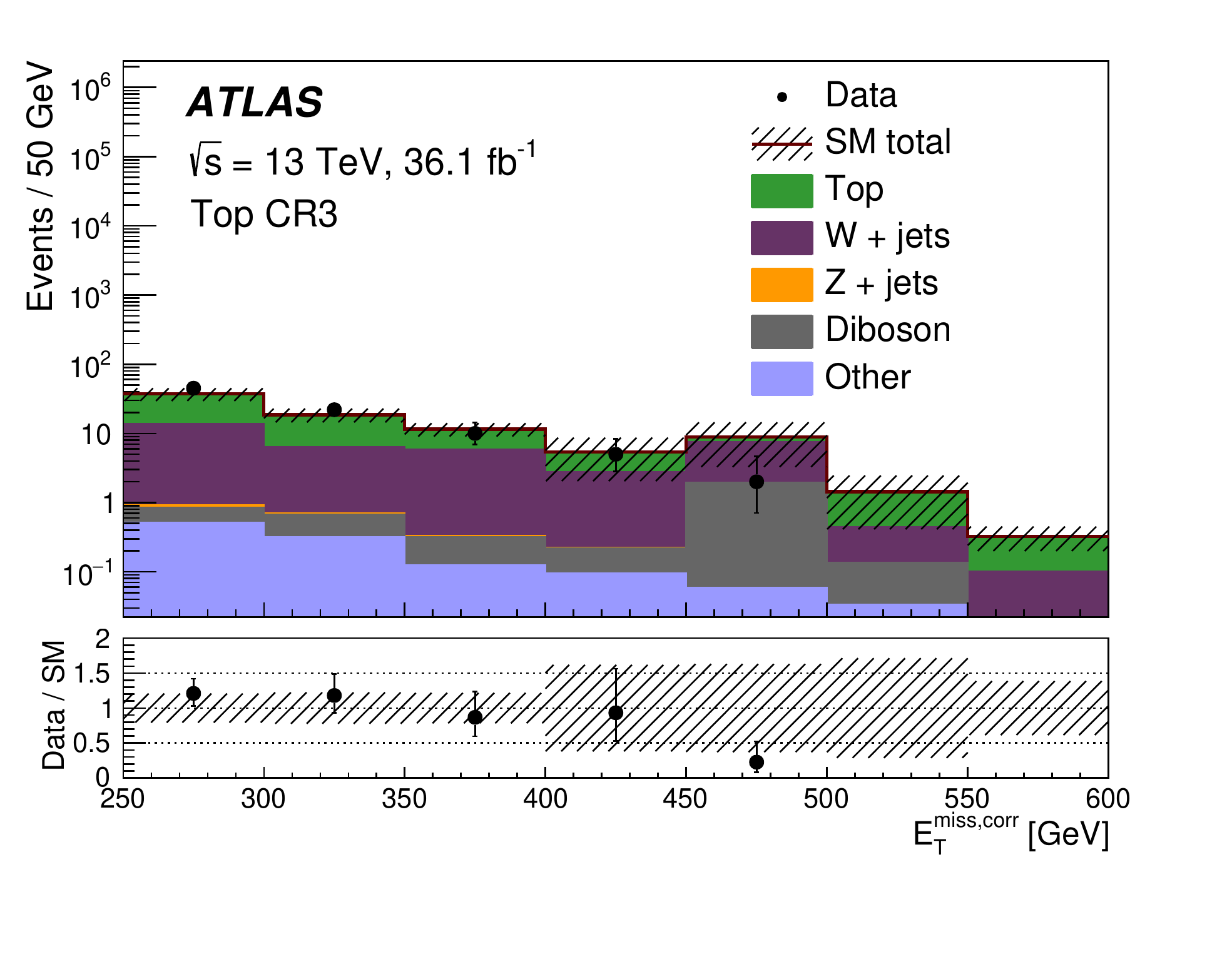}
   }
   \subfloat[\label{fig:TopCR5}]{
    \includegraphics[width=0.49\textwidth]{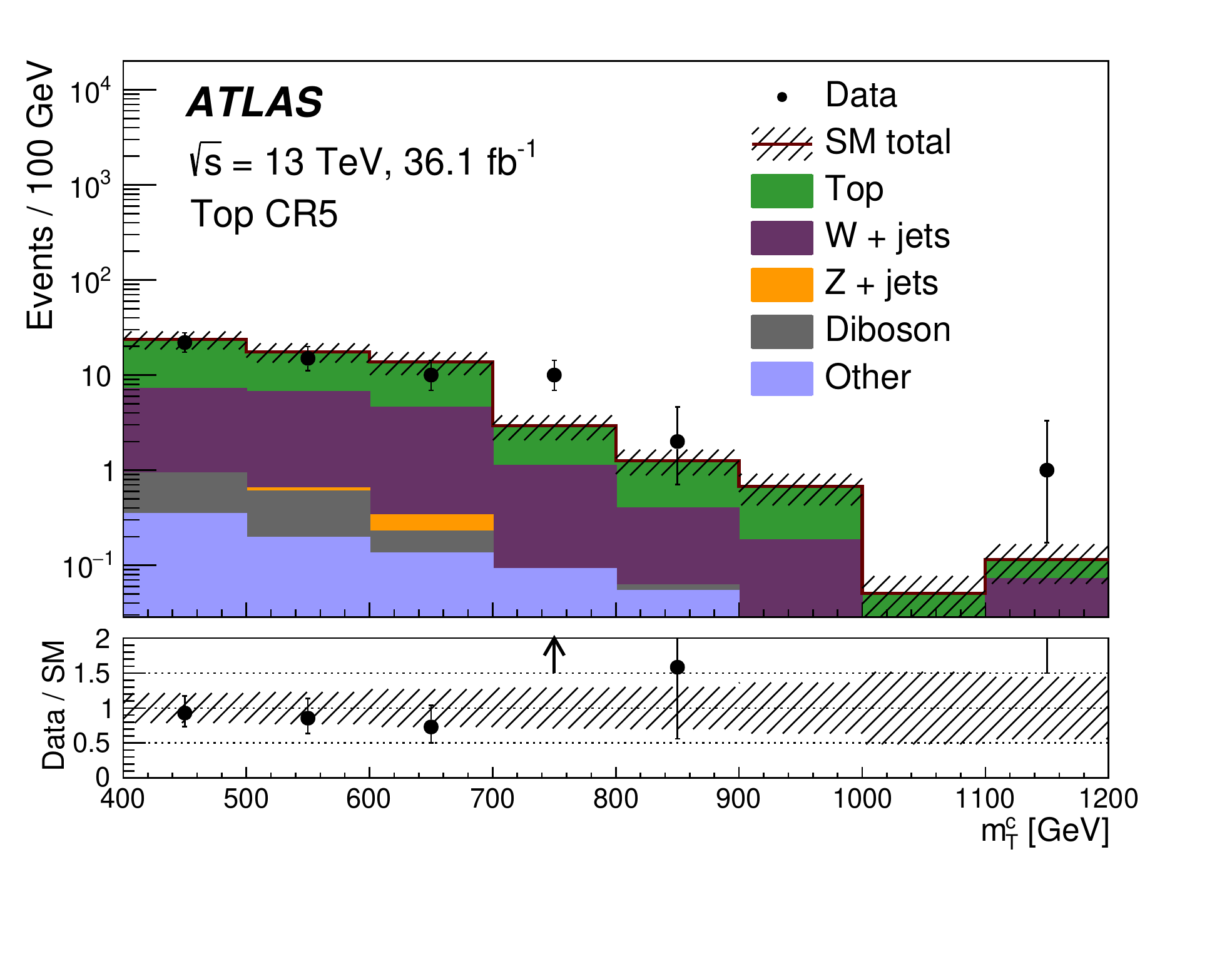}
   }
   \caption{ Examples of comparisons between data and expectation after the background-only fit for the \metcorr{} distribution \protect\subref{fig:ZCR1}  \Z{} CR1, \protect\subref{fig:WCR2}  \W{} CR2  and  \protect\subref{fig:TopCR3} Top CR3, and for the \mtc{} distribution \protect\subref{fig:ZCR5} \Z{} CR5, \protect\subref{fig:WCR4}  \W{} CR4 and \protect\subref{fig:TopCR5} Top CR5 . The contributions from all SM backgrounds are shown as a histogram stack. The shaded band indicates the detector-related systematic uncertainties and the statistical uncertainties of the MC samples. The error bars on the data points indicate the data's statistical uncertainty. The lower panel shows the ratio of the data to the SM prediction after the background-only fit.
   }
   \label{fig:CR}
 \end{figure}

Figure \ref{fig:CR} shows examples of comparisons between data and the SM prediction for the \metcorr{} and \mtc{} distributions, for some of the \Z{} CRs, \W{} CRs and Top CRs. 
The backgrounds are scaled by the normalisation factors extracted from the background-only fit. 
Good agreement between data and the prediction is observed in the shapes of all distributions.

\subsection{Validation regions}
\label{sec:VR}

After the background-only fit is performed, the validity of the extrapolation of the background estimate outside the control regions is tested in dedicated validation regions (VRs).
Validation regions are not used to constrain the background normalisation in the fit. They are required to have kinematic properties and flavour-tagging compositions similar to those in the signal regions, while keeping the potential contamination from signal as low as possible. 
This is achieved by retaining most of the selection criteria used in the signal region and inverting or restricting the requirements on \met{}, \mtc{}, and the jet momenta to limit the potential signal contribution. All validation region selections require $\mjjw>$~\SI{125}{\GeV} and $\mtc>$~\SI{120}{\GeV}.
Depending on the signal region, two or three VRs (referred to as A, B and C) are defined to separately validate the modelling of kinematic variables used in the signal region definitions: \met{}, \jetOnePt{} and \jetTwoPt{}. 
A full list of all the selection criteria used to define the validation regions is reported in Table \ref{tab:VR}. The signal contamination in the validation regions was examined for all considered signals that have not yet been excluded, and was found to be at most 25\%. 

\begin{figure}[h]\centering
     \includegraphics[width=0.80\textwidth]{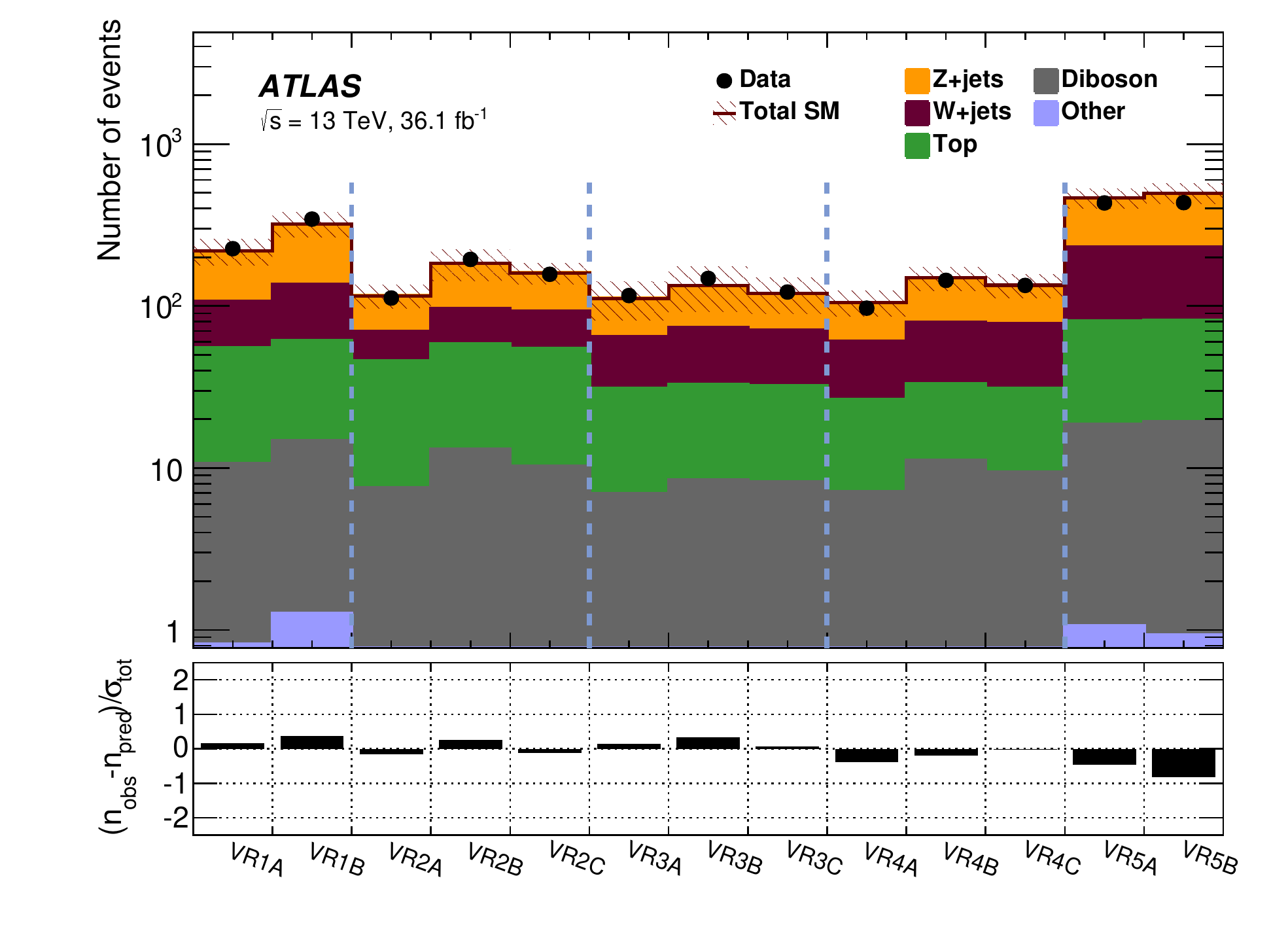}   
     \caption{Comparison of the numbers of data events and predicted SM yields in each validation region. The normalisation of the backgrounds is obtained from the fit to the CRs. The shaded band indicates the total background uncertainty. The lower panel shows the pulls, estimated as the difference between the observed number of events ($n_{\mathrm obs}$) and the predicted background yields ($n_{\mathrm pred}$) divided by the total uncertainty ($\sigma_{\mathrm tot}$). }
   \label{fig:pullVR}
 \end{figure}

Figure \ref{fig:pullVR} shows a comparison between the number of events predicted by the background-only fit and the data, for all validation regions. The pulls, estimated as the difference between the observed number of events and the predicted background yields divided by the total uncertainty, are also shown. No evidence of significant background mismodelling is observed in the VRs.

\begin{table}[th]
  \begin{centering}
    \renewcommand{\arraystretch}{1.1}
    \begin{tabular}{l|c|c|c|c|c}  
      & VR1 & VR2 & VR3 & VR4 & VR5 \\
      \hline
      \hline
      Trigger & \multicolumn{5}{c}{\met{} triggers} \\
      Leptons & \multicolumn{5}{c}{0 $e$ \texttt{AND} 0 $\mu$ } \\
      \met{} [GeV]           & \multicolumn{5}{c}{$ > 500$}\\
      \DphiJetMet{} [rad]            &   \multicolumn{5}{c}{$> 0.4$}\\

      \nctag{} & \multicolumn{5}{c}{ $\geq 1$ }\\
      \mjjw{} [GeV]             & \multicolumn{5}{c}{$ > 125$}\\
      \hline
      \njets{}                 & $\geq 2$     & $\geq 3$ & $\geq 3$         & $\geq3$   & $\geq 3$ \\
      Leading jet $c$-tag veto & yes          & yes      & yes              & yes       & no      \\
      \jetOnePt{} [GeV]      &   $>250$  &  $>250$  &  $>250$  & $>250$   & $>300$   \\
      \cTagOnePt{} [GeV]     & $<100$       & $>60$    & $>80$            & $>100$    & $>200$   \\
      \mtc{} [GeV]           & $>250$             & $\in(300, 450)$               & $\in(300,400)$       & $\in(300,500)$    & $>400$ \\
      \hline
      \hline
      \multirow{2}*{A}\ \ \ \ \ \  \met{} [GeV] &  $<350$       & $<350$       & $<350$       & $<350$       & $<350$ \\ 
                        \ \ \ \ \ \ \ \ \ \jetTwoPt{} [GeV]   &     $-$             & $<175$  & $<200$  & $<200$  & $-$ \\
      \hline
      \multirow{2}*{B}\ \ \ \ \ \  \jetOnePt{} [GeV]   &  $<350$  & $<350$  & $<350$  & $<350$  &   $-$  \\
                              \ \ \ \ \ \ \ \ \ \jetTwoPt{} [GeV]   &       $-$           & $<175$  & $<200$  & $<200$  & $<175$    \\
      \hline
      \multirow{2}*{C} \ \ \ \ \  \met{} [GeV]   &        $-$          & $<350$       & $<350$       & $<350$       & $-$\\
                                \ \ \ \ \ \ \ \ \ \jetOnePt{} [GeV]   &         $-$         & $<350$  & $<350$  & $<350$  & $-$\\
    \end{tabular}
      \par\end{centering}{\tiny \par}
     \caption{\label{tab:VR} 
     Overview of the validation region selection criteria. 
     \njets{} and \nctag{} indicate the total number of jets and \cjets{}, respectively; 
     \jetOnePt{} and \jetTwoPt{} indicate the transverse momentum of the leading and sub-leading jet; and  
     \cTagOnePt{} is the transverse momentum of the \cjet{} with the highest \pt{}. 
     }
\end{table}

\clearpage

%% file: latex/sections/SystematicsUncertainties.tex
Various sources of systematic uncertainty are considered in this analysis. 
Uncertainties are treated as correlated across different regions where appropriate. 
Uncertainties corresponding to the number of events in the control regions used to constrain background normalisations are also considered. 

The dominant experimental systematic effects on the estimated background rates are due to the uncertainties in the $c$-tagging efficiency  and the jet energy scale (JES)~\cite{PERF-2016-04}. The flavour-tagging uncertainties are estimated by varying the $\eta$-, \pt- and flavour-dependent scale factors applied to each jet in the simulation within a range that reflects the systematic uncertainty in the measured tagging efficiency and mis-tag rates in 13~\TeV\ data. This uncertainty in the scale factors is in the range 15--40\%, 3--12\%, and 11--18\% for the light-flavour jet, $b$-jet and $c$-jet scale factors, respectively. The uncertainties on the estimated background rates  due to the uncertainties in the $c$-tagging efficiency and the JES are found to be 7$-$9\% and 5$-$8\%, respectively, across the different signal regions. 
The uncertainties associated with lepton reconstruction and energy measurements, as well as the jet energy resolution~\cite{PERF-2011-04}, are also considered but have a negligible impact on the final results. Lepton and jet-related uncertainties are propagated to the \met{} calculation, and additional uncertainties are included in the energy scale and resolution of the soft term \cite{ATL-PHYS-PUB-2015-023}. 

Theoretical and modelling uncertainties of the top-quark-pair background are computed as the difference between the prediction from nominal samples and those from alternative samples differing in parameter settings or in the choice of generator. Hadronisation and PS uncertainties are estimated using samples generated using \POWHEGBOX v2 and showered by \herwigpp v2.7.1~\cite{Bahr:2008pv} with the UEEE5~\cite{Gieseke:2012ft}  underlying-event tune. Uncertainties related to initial- and final-state radiation modelling, PS tune and choice of the $h^{}_{\mathrm{damp}}$ parameter value in \POWHEGBOX v2 are estimated using alternative settings of the event generators. 
Finally, an alternative sample produced using the \SHERPA 2.2.1 generator with settings as those defined for $V$+jets samples, is used to estimate the event generator uncertainties.

Theoretical uncertainties in the modelling of $V$+jets production are assessed by considering alternative samples with varied merging (CKKW-L), resummation, renormalisation, and factorisation scales, as well as PDF variations from the NNPDF3.0NNLO replicas.
The uncertainties in the background estimates in the SRs arising from renormalisation scale and PDF variations are 6$-$11\% and 7$-$15\%, respectively. In all SRs, they are among the dominant systematic uncertainties. 

To account for differences in the heavy-flavour composition between the \W{} CR4 and SR4, an additional normalisation uncertainty of 54\% \cite{STDM-2011-21} is assigned in these regions to the heavy-flavour component of $W$+jets. The effect of this uncertainty on the background estimate is found to be small, around 1\%. 

Uncertainties in the diboson background are evaluated by considering different resummation, normalisation, and factorisation scales using alternative samples. 
An additional 15\% uncertainty is assigned to its normalisation, accounting for the specific choice of the EWK parameters. 
The single-top-quark background forms less than $2\%$ of the total background in all signal regions; a 100\% uncertainty is assigned to its normalisation, with no impact on the search sensitivity.

The dominant relative uncertainties affecting the background estimates in each signal region are summarised in Table~\ref{tab:syst_summary}. The uncertainties on the $\mu$ factors are also included. 

\begin{table}[htbp]
\begin{center}
\setlength{\tabcolsep}{0.0pc}
{\small
  \begin{tabular*}{\textwidth}{@{\extracolsep{\fill}}l rrrrr}
    \noalign{\smallskip}\hline\noalign{\smallskip}
  \textbf{Source \textbackslash~Region}    &     SR1    & SR2 & SR3 &  SR4 & SR5 \\[-0.05cm]
      &     [\%]    & [\%] & [\%] &  [\%] & [\%] \\
     \noalign{\smallskip}\hline\noalign{\smallskip}
  \midrule
  $\mu_Z$ & 6.7 & 9.3 & 12 & 11 & 13\\
  $\mu_W$ & 4.5 & 5.6 & 4.8 & 4.4 & 3.9\\
  $\mu_\mathrm{Top}$ & 2.9 & 7.7 & 2.2 & 2.0 & 2.1\\
  \midrule
  JES  & 7.9 & 5.0 & 6.8 & 5.2 & 5.6\\
  $c$-tagging & 6.7 & 8.9 & 9.3 & 8.1 & 7.0\\
  \midrule 
  $W/Z$+jets scale variations & 11 & 5.8 & 7.6 & 6.5 & 5.2\\
  $W/Z$+jets resummation scale & 7.8 & 3.7 & 2.5 & 5.6 & 4.7\\
  $W/Z$+jets PDF & 7.7 & 7.1 & 14 & 15 & 9.1\\
  \midrule
  Total &  18 & 15 & 19 & 18 & 16\\
    \noalign{\smallskip}\hline\noalign{\smallskip}
    \end{tabular*}
     }
\caption{\label{tab:syst_summary} Summary of the dominant experimental and theoretical uncertainties for each signal region.
Uncertainties are quoted as percentages of the total SM background predictions.
The individual uncertainties can be correlated, and do not necessarily add in quadrature to the total background uncertainty.}
     \end{center}
     \vspace{-0.2cm}
\end{table}

For the SUSY signal processes, both the experimental and theoretical uncertainties in the expected signal yield are considered. Experimental uncertainties are found to be between 11\% and 24\% across the $\stopone,\scharmone$--$\ninoone$ mass plane. 
In all SRs, they are largely dominated by the uncertainty in the $c$-tagging efficiency. They vary between 10\% and 20\% for squark masses in the range between 300~\GeV\ and 1000~\GeV. Additional uncertainties in the acceptance and efficiency of the SUSY signal MC samples are considered, due to the variation of the QCD coupling constant $\alphas$, the renormalisation and factorisation scales, the CKKW scale used to match the parton shower and matrix element descriptions, and the parton shower tunes.  The total magnitude of these uncertainties ranges up to about 13\%, depending on the signal region and sparticle masses considered. 

Theoretical uncertainties in the NLO+NLL cross-section are calculated for each SUSY signal scenario and are dominated by the uncertainties in the renormalisation and factorisation scales, followed by the uncertainty in the PDF. 
These uncertainties are not treated using a nuisance parameter in any of the fits considered, since their impact is quantified separately.

%% file: latex/sections/Results.tex
\begin{figure}[tbh]\centering
     \includegraphics[width=0.80\textwidth]{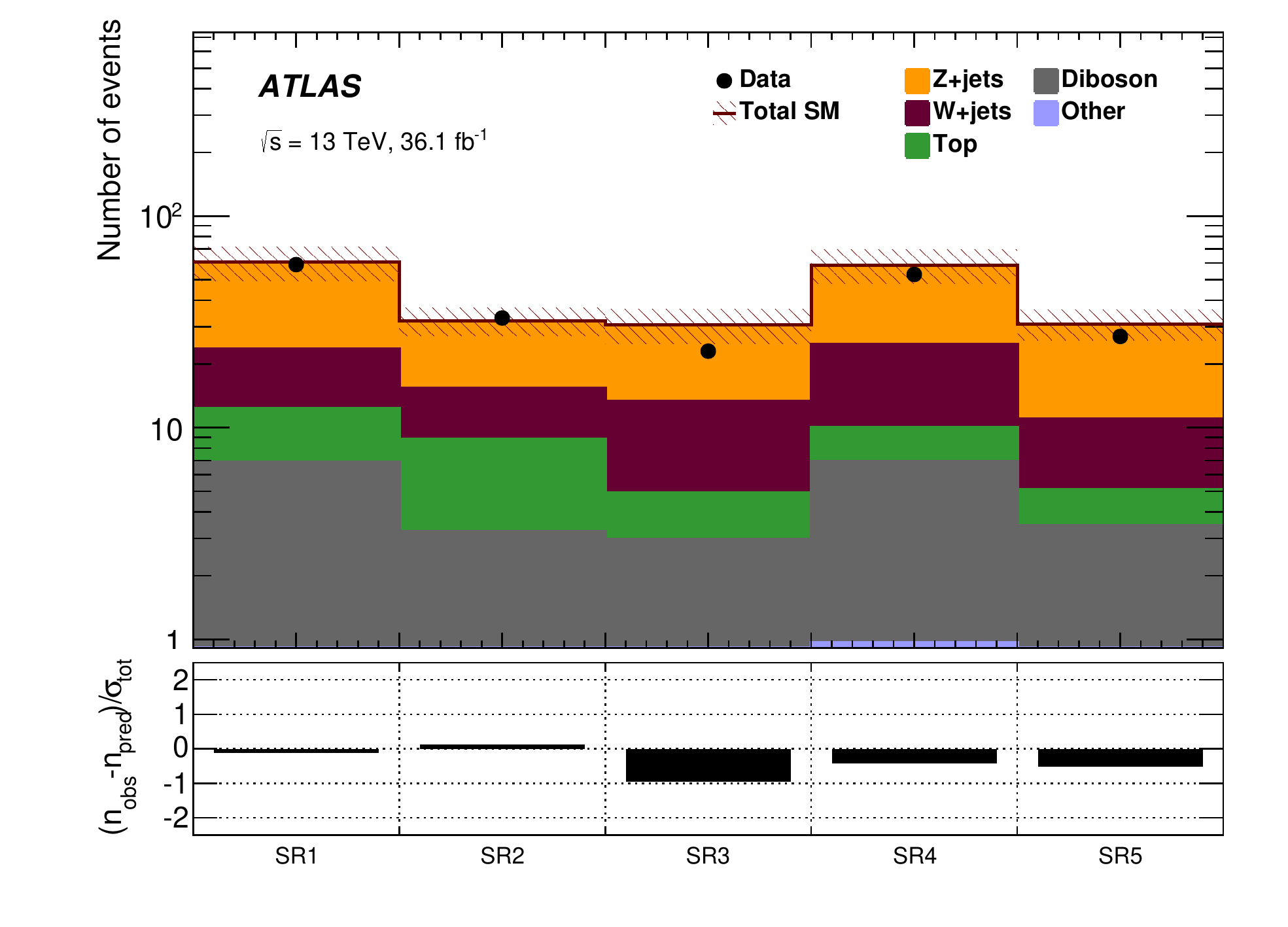}   
     \caption{ Comparison of the numbers of data events and predicted SM yields in each signal region. The normalisation of the backgrounds is obtained from the fit to the CRs. The shaded band indicates the total background uncertainty. The lower panel shows the pulls, estimated as the difference between the observed number of events ($n_{\mathrm obs}$) and the predicted background yields ($n_{\mathrm pred}$) divided by the total uncertainty ($\sigma_{\mathrm tot}$).
}
   \label{fig:pullSR}
 \end{figure}

The observed event yields for each signal region are shown in Table~\ref{tab:SRyields} together with the estimated background contributions after the fit. The total uncertainty in the estimated background ranges from 15\% to 20\%, depending on the signal region, and is dominated by the uncertainty in the $Z$+jets background.
No significant excess above the SM prediction obtained after the fit is seen in any of the regions. The smallest observed $p$-value, which expresses the probability that the background fluctuates to the data or above, is 0.41 in region SR2. The largest deficit is observed in SR3, where 23 events are observed while $31\pm 6$ events are predicted. Figure \ref{fig:pullSR} shows a comparison of the observed and predicted yields in all SRs, together with the pull in each region.

\begin{figure}[]\centering
   \subfloat[\label{fig:SR1}]{
     \includegraphics[width=0.49\textwidth]{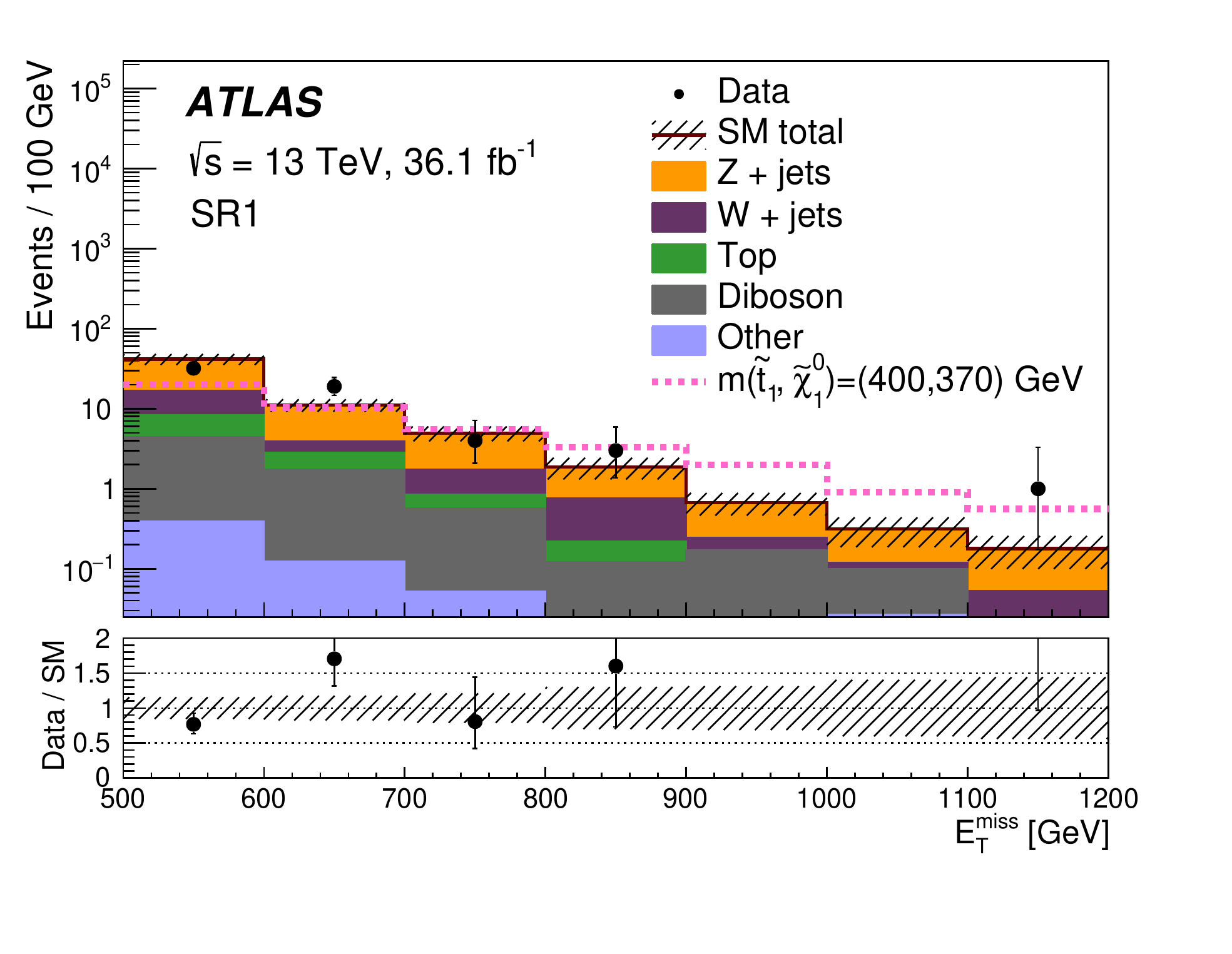}
   }
   \subfloat[\label{fig:SR2}]{
    \includegraphics[width=0.49\textwidth]{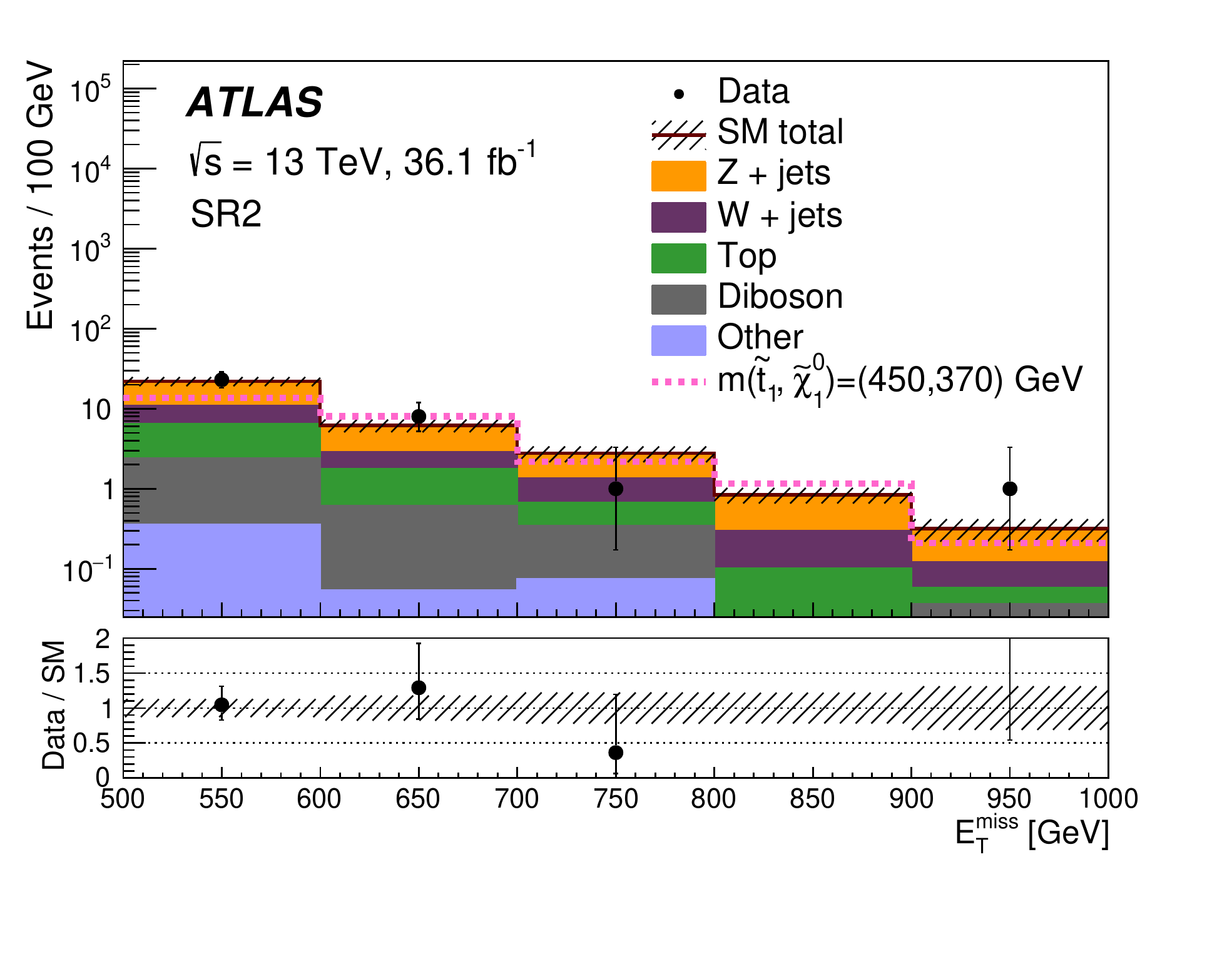}
   }\\
   \subfloat[\label{fig:SR3}]{
    \includegraphics[width=0.49\textwidth]{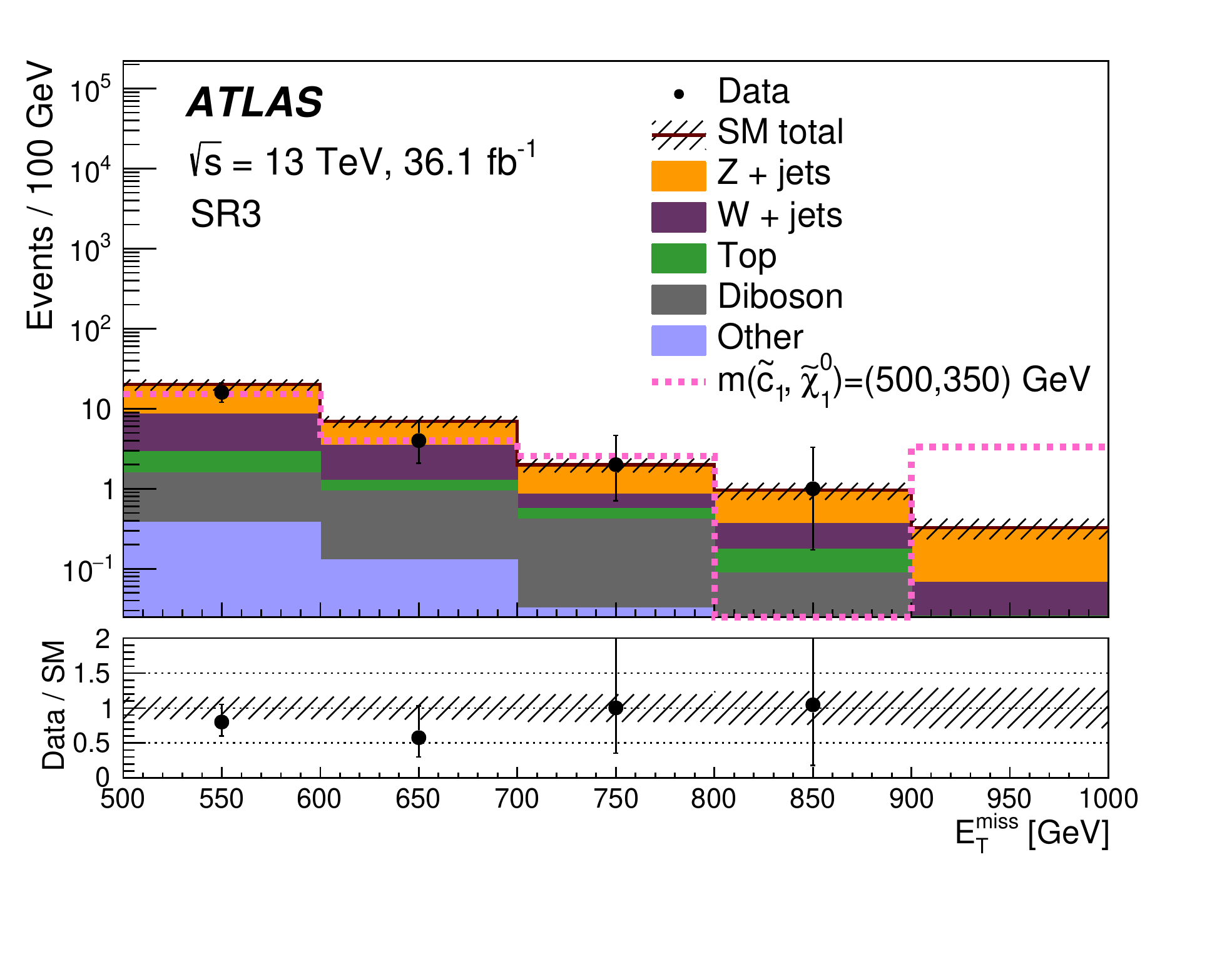}
   }
   \subfloat[\label{fig:SR4}]{
    \includegraphics[width=0.49\textwidth]{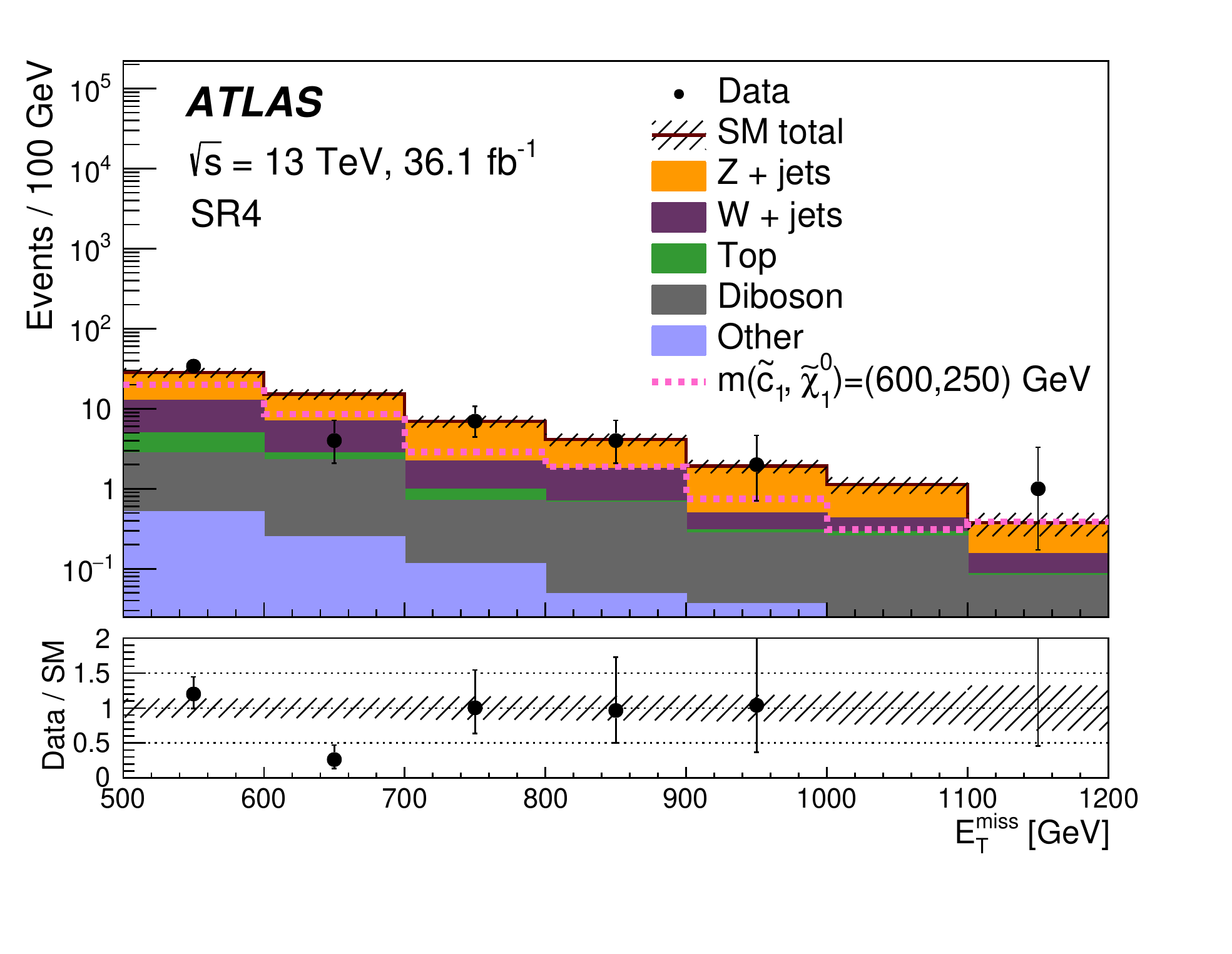}
   }\\
   \subfloat[\label{fig:SR5}]{
    \includegraphics[width=0.49\textwidth]{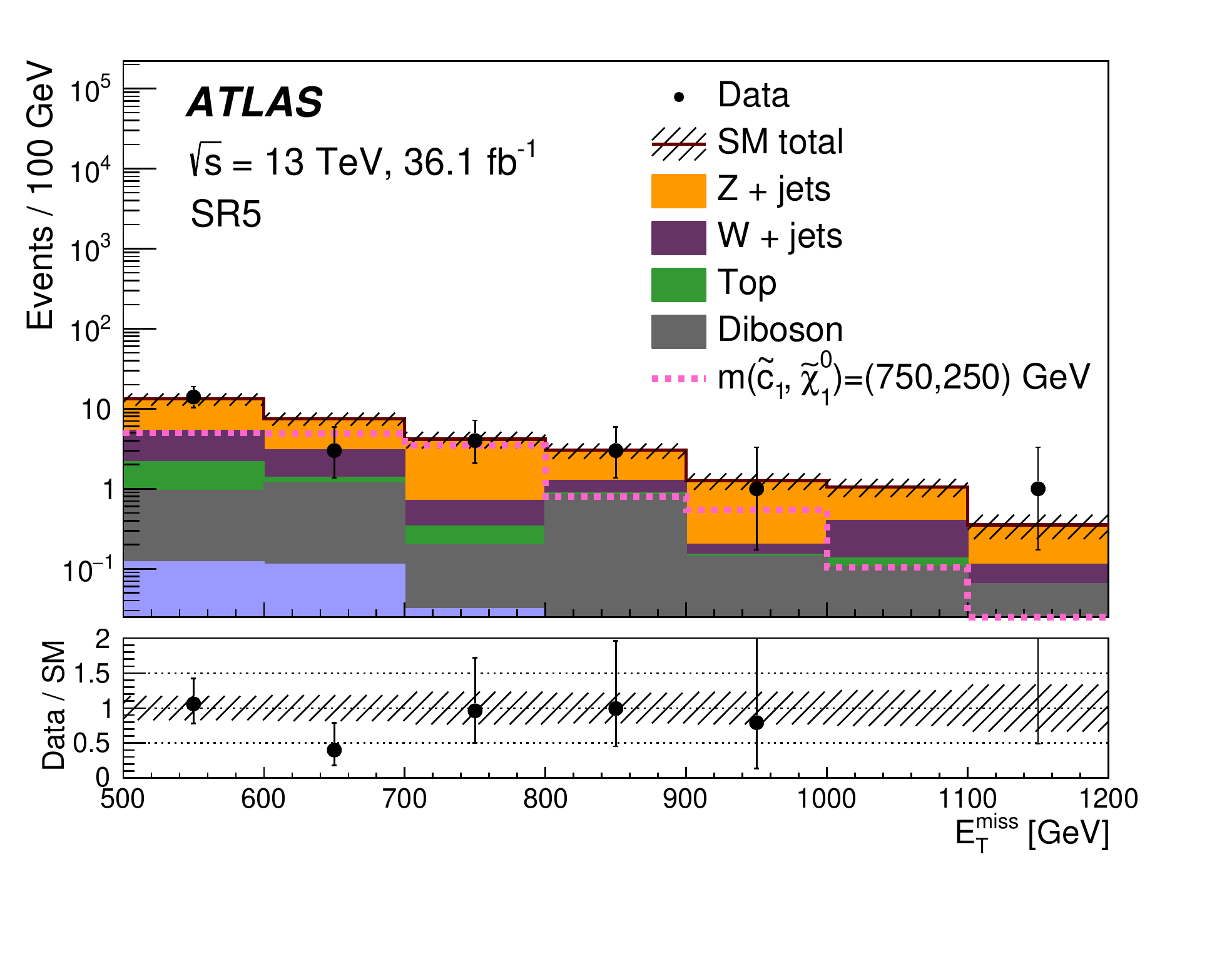}
   }
   \caption{ Comparison between data and expectation after the background-only fit  for the \met{} distribution \protect\subref{fig:SR1} in SR1, \protect\subref{fig:SR2} in SR2, \protect\subref{fig:SR3} in SR3, \protect\subref{fig:SR4} in SR4, and \protect\subref{fig:SR5} in SR5. The shaded band indicates the detector-related systematic uncertainties and the statistical uncertainties of the MC samples, while the error bars on the data points indicate the data's statistical uncertainty. The final bin in each histogram includes the overflow. The lower panel shows the ratio of the data to the SM prediction after the background-only fit. In each plot, the distribution is also shown for a representative signal point.}
   \label{fig:SR}
 \end{figure}

Comparisons of the data to the SM predictions are shown in Figure \ref{fig:SR} for the \met{} distribution in each SR. The expected distribution from a representative signal point is overlaid on each plot. Good agreement in the shapes of the data and predicted distributions is observed in all cases and again no significant excess above the SM prediction is observed.

\begin{table} 
\begin{center} 
\renewcommand{\arraystretch}{1.2}
\setlength{\tabcolsep}{0.0pc} 
{\footnotesize
\begin{tabular*}{\textwidth}{@{\extracolsep{\fill}}lrrrrr} 

\noalign{\smallskip}\hline\noalign{\smallskip}
   \textbf{Yields}            &  SR1               &  SR2               &  SR3               &  SR4               &   SR5               \\ 
        \noalign{\smallskip}\hline\noalign{\smallskip} 
Observed           &  $59$                     &  $33$                     &  $23$                     &  $53$                     &   $27$            \
         \\
     \noalign{\smallskip}\hline\noalign{\smallskip}
     Total SM          &  $61 \pm 11$         & $32\pm5$ & $31\pm6$ & $59\pm11$ & $31\pm5$\\
     \noalign{\smallskip}\hline\noalign{\smallskip}
        $Z$+jets      & $37.1 \pm 7.8$ & $16.7\pm3.2$ & $17\pm5$ & $34\pm8$ & $20\pm4$\\
        $W$+jets      & $11.2 \pm 5.1$ & $6.5\pm2.3$ & $8.4\pm2.0$ & $15\pm4$ & $5.9\pm1.5$\\
        Top           & $5.4 \pm 2.0$ & $5.6\pm2.6$ & $2.0\pm 2.0$ & $3.1\pm1.8$ & $1.7\pm0.7$\\
        Diboson      &   $6.3 \pm 2.1$ & $2.7\pm1.7$ & $2.4\pm0.7$ & $5.9\pm2.3$ & $3.2\pm1.6$\\
        Other    & $0.6 \pm 0.1$ & $0.5\pm0.1$ & $0.5\pm0.1$ & $1.0\pm0.1$ & $0.3\pm0.1$\\ 
    \noalign{\smallskip}\hline\noalign{\smallskip} 
	Signal benchmarks &                 &                 &                 &                 &                  \\ 
    \noalign{\smallskip}\hline\noalign{\smallskip} 
	$(m_{\stopone},m_{\ninoone}) =(450,425)$ GeV &  $22.7\pm4.0$  &  $9.1\pm2.6$   & $1.6\pm1.0$    &  $1.84\pm0.71$ &   $0.45\pm0.27$    \\ 
	$(m_{\stopone},m_{\ninoone}) =(500,420)$ GeV &  $18.3\pm3.4$  &  $19.7\pm4.9$  & $15.2\pm4.1$   &  $8.0\pm2.2$   &   $1.26\pm0.64$    \\ 
	$(m_{\stopone},m_{\ninoone}) =(500,350)$ GeV &  $5.4\pm2.0$   &  $11.6\pm3.3$  & $26.1\pm6.7$   &  $18.7\pm5.4$  &   $3.0\pm1.1$      \\ 
	$(m_{\stopone},m_{\ninoone}) =(600,350)$ GeV &  $1.91\pm0.87$ &  $3.2\pm1.3$   & $10.5\pm3.0$   &  $24.0\pm5.9$  &   $7.0\pm2.2$      \\ 
	$(m_{\stopone},m_{\ninoone}) =(900,1)$ GeV &    $0.67\pm0.19$ &  $0.61\pm0.21$ & $1.61\pm0.50$  &  $11.7\pm2.0$  &   $10.2\pm1.8$   \\ 
    \noalign{\smallskip}\hline\noalign{\smallskip} 

$\langle \sigma_\textrm{vis} \rangle_{\mathrm obs}^{95}$ [fb]  & $0.67$ & $0.46$ & $0.33$ & $0.59$ & $0.40$ \\
$S_{\mathrm obs}^{95}$  & $24.2$  & $16.6$ &  $11.9$ & $21.3$ & $14.3$ \\
$S_{\mathrm exp}^{95}$ &  $ { 24.4 }^{ +13.2 }_{ -7.6 }$ & $ { 16.0 }^{ +5.6 }_{ -4.4 }$ & $ { 15.0 }^{ +5.2 }_{ -3.1 }$ & $ { 24.9 }^{ +9.6 }_{ -7.1 }$ & $ { 15.3 }^{ +6.8 }_{ -2.2 }$ \\
$p(s=0)$ &   $0.5$ & $0.41$ & $0.5$ & $0.5$ & $0.5$ \\
\noalign{\smallskip}\hline\noalign{\smallskip} 
\end{tabular*} 
} 
\end{center} 
\caption{Observed and expected yields for all the signal regions considered in the analysis. The uncertainties include experimental, theoretical, and Monte Carlo statistical uncertainties. Benchmark signal models yields are also given for each SR.  
The lower part of the table reports the 95\% CL upper limits on the visible cross-section, $\langle \sigma_\textrm{vis} \rangle_{\mathrm obs}^{95}$, on the number of signal events, $S_{\mathrm obs}^{95}$, and on the number of signal events given the expected number (and $\pm 1\sigma$ excursions of the expectation) of background events, $S_{\mathrm exp}^{95}$. The discovery $p$-value, $p(s=0)$, where $s$ is the number of signal events, is also reported. The value of 0.5 is assigned in the case of a downward data fluctuation where the number of observed events is less than the expected number of events.  
} 
\label{tab:SRyields} 
\end{table} 

In the absence of a signal, expected and observed 95\% confidence level (CL) limits ($S^{95}_\mathrm{exp}$ and $S^{95}_\mathrm{obs}$, respectively) are set on the number of possible events from sources beyond the Standard Model (BSM) in each signal region using the CL$_\text{s}$ prescription~\cite{Junk:1999kv,Read:2002hq}. Limits on the visible cross-section $\sigma^{}_\textrm{vis}$ of a BSM signal, defined as the product of the production cross-section $\sigma$, the acceptance $A$ and the selection efficiency $\epsilon$, are also computed as $\sigma^{}_\textrm{vis} = S^{95}_\mathrm{obs}/\int {\cal{L}}{\mathrm{d}}t$, where $\int {\cal{L}}{\mathrm{d}}t$ is the integrated luminosity. The results are shown in the lower part of Table~\ref{tab:SRyields}.

\begin{figure}[t]
\begin{center}
 \includegraphics[width=0.70\textwidth]{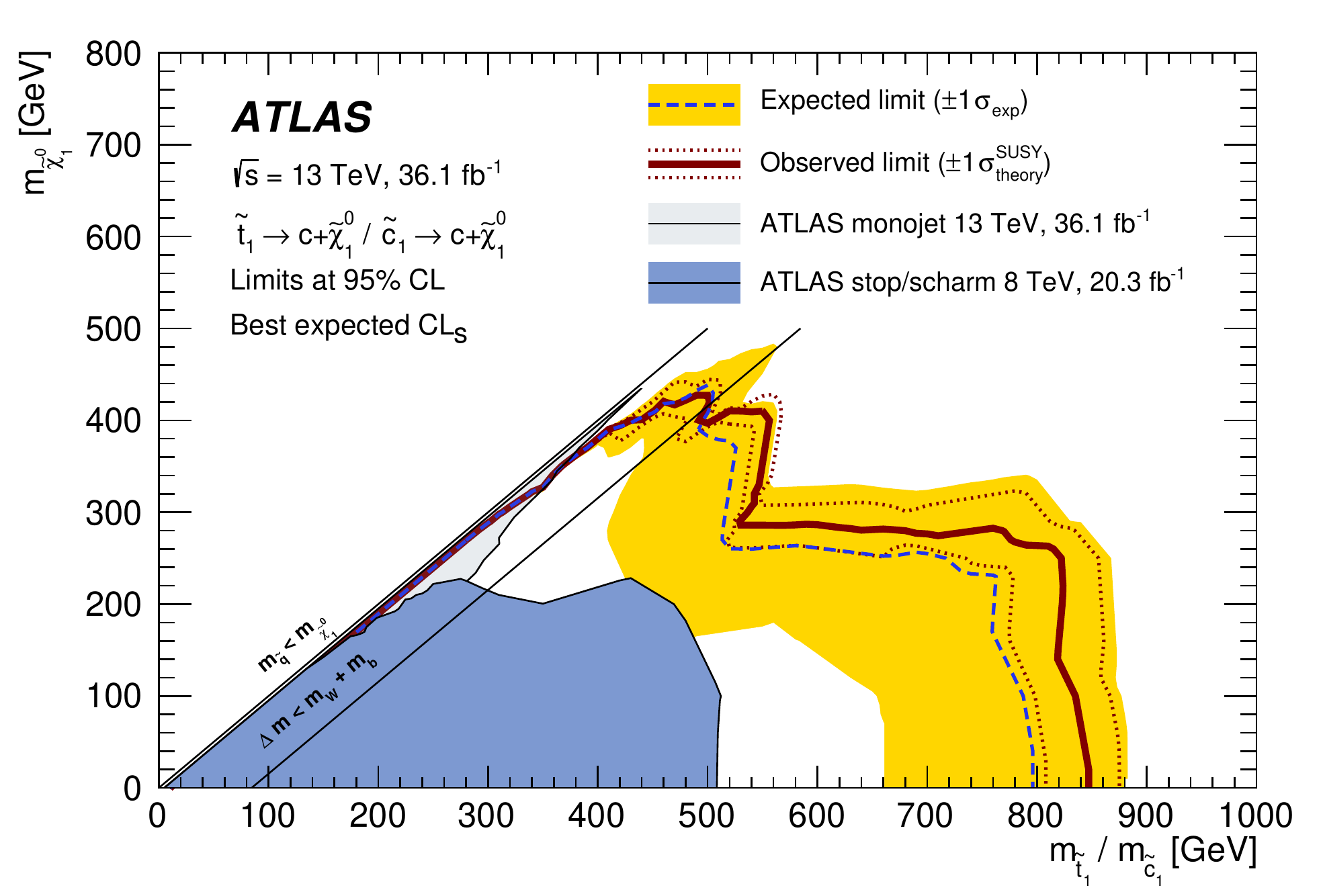} 
\end{center}
\caption{ \label{fig:limits} 
Observed and expected exclusion contours at 95\% CL, as well as $\pm 1 \sigma$ variation of the expected limit, in the $\stopone, \scharmone$--$\ninoone$ mass plane. The SR with the best expected sensitivity is adopted for each point of the parameter space. The solid band around the expected limit (dashed line) shows the impact of the experimental and SM background theoretical uncertainties. The  dotted lines show the impact on the observed limit of the variation of the nominal signal cross-section by $\pm 1 \sigma$ of its theoretical uncertainties. 
The observed limit from the monojet search at $\sqrt{s}=$ \SI{13}{TeV} \cite{Aaboud:2017phn} and from the Run-1 analysis \cite{Aad:2014nra,Aad:2015gna} at $\sqrt{s}=$ \SI{8}{TeV} are also overlaid. $\Delta m$ denotes $m_{\stopone, \scharmone}-m_{\ninoone}.$
}
\end{figure}

\begin{figure}[h] 
  \begin{center}
   \includegraphics[width=0.7\textwidth]{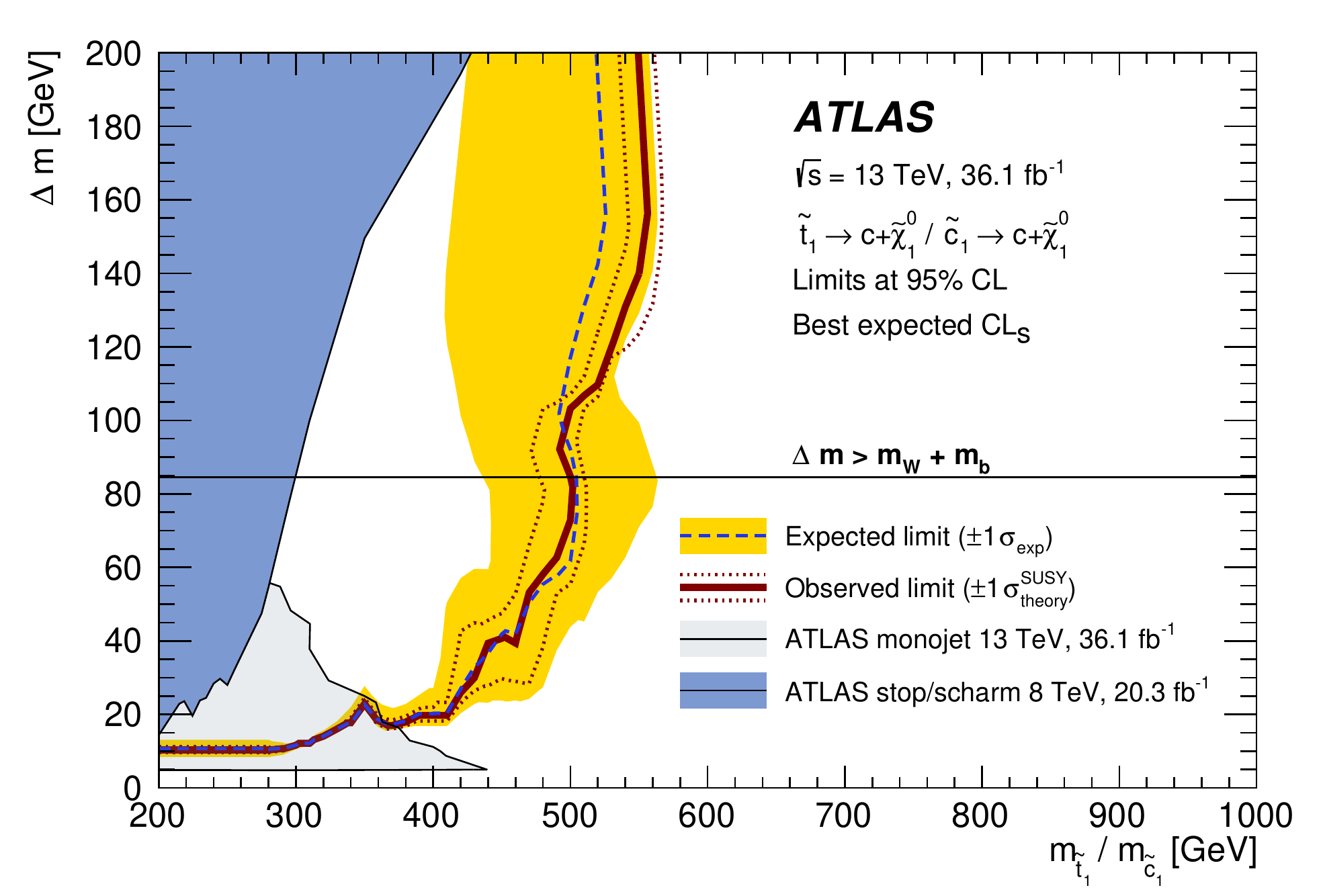}
    \caption{
    Observed and expected exclusion contours at 95\% CL, as well as $\pm 1 \sigma$ variation of the expected limit, in the $m_{\stopone, \scharmone}$--$\Delta m$ plane, where $\Delta m$ denotes $m_{\stopone, \scharmone}-m_{\ninoone}$. The SR with the best expected sensitivity is adopted for each point of the parameter space. The solid band around the expected limit (dashed line) shows the impact of the experimental and SM background theoretical uncertainties. The  dotted lines show the impact on the observed limit of the variation of the nominal signal cross-section by $\pm 1 \sigma$ of its theoretical uncertainties. 
The observed limit from the monojet search at $\sqrt{s}=$ \SI{13}{TeV} \cite{Aaboud:2017phn} and from the Run-1 analysis \cite{Aad:2014nra,Aad:2015gna} at $\sqrt{s}=$ \SI{8}{TeV} are also overlaid. }
    \label{fig:limits2}
  \end{center}
\end{figure}

The profile-likelihood-ratio test statistic is used to set limits on the top-squark and charm-squark pair production signal models considered in this paper. If the observed CL$_\text{s}$ value for a particular signal model is smaller than 0.05, the model is excluded at the 95\% CL. The results are presented in the $\stopone, \scharmone$--$\ninoone$ mass plane in Figure~\ref{fig:limits}. At each point in the mass plane, the signal region with the smallest expected value of CL$_\text{s}$ is used to set the limit. 
Changes in the trend of the exclusion contour are observable around m($\stopone, \scharmone$,$\ninoone$)~=~(480, 400)~GeV and m($\stopone, \scharmone$,$\ninoone$)~=~(500, 300)~GeV. They are due to the transitions between SR2 and SR3, and between SR3 and SR4, respectively. 
The grey region has been excluded by ATLAS searches with an energetic jet and large missing transverse momentum described in Ref.~\cite{Aaboud:2017phn}. This region is characterised by events with soft $c$-jets, for which the efficiency of the signal regions of the present analysis is very low. For larger values of $\Delta m$, the search in Ref.~\cite{Aaboud:2017phn} loses sensitivity due to a veto on jet activity beyond four jets, required in the signal regions of the search. 
For signal models characterised by $\Delta m$~<\SI{200}{\GeV} the same results are also presented in the $m_{\stopone, \scharmone}$--$\Delta m$ plane as shown in Figure~\ref{fig:limits2}. 

%% file: latex/sections/Conclusion.tex
This paper reports a search for direct top-squark and charm-squark pair production in final states with charm-tagged jets and missing transverse momentum, based on a \SI{36.1}{\per\femto\barn} dataset of $\sqrt{s} =$ \SI{13}{\TeV} proton--proton collisions recorded by the ATLAS experiment at the LHC in 2015 and 2016. 
No significant excess above the expected Standard Model background is found and model-independent exclusion limits at 95\% CL are set on the visible cross-section. 
Assuming a 100\% branching ratio to $c\ninoone$, top and charm squarks with masses up to \SI{850}{\GeV} are excluded at 95\% CL for a massless lightest neutralino. For  $m_{\stopone,\scharmone}-\mn <$~\SI{100}{\GeV}, top and charm squarks with masses up to \SI{500}{\GeV} are excluded.

%% file: acknowledgements/Acknowledgements.tex
% Acknowledgements for papers with collision data
% Version 14-Feb-2018

% Standard acknowledgements start here
%----------------------------------------------
We thank CERN for the very successful operation of the LHC, as well as the
support staff from our institutions without whom ATLAS could not be
operated efficiently.

We acknowledge the support of ANPCyT, Argentina; YerPhI, Armenia; ARC, Australia; BMWFW and FWF, Austria; ANAS, Azerbaijan; SSTC, Belarus; CNPq and FAPESP, Brazil; NSERC, NRC and CFI, Canada; CERN; CONICYT, Chile; CAS, MOST and NSFC, China; COLCIENCIAS, Colombia; MSMT CR, MPO CR and VSC CR, Czech Republic; DNRF and DNSRC, Denmark; IN2P3-CNRS, CEA-DRF/IRFU, France; SRNSFG, Georgia; BMBF, HGF, and MPG, Germany; GSRT, Greece; RGC, Hong Kong SAR, China; ISF, I-CORE and Benoziyo Center, Israel; INFN, Italy; MEXT and JSPS, Japan; CNRST, Morocco; NWO, Netherlands; RCN, Norway; MNiSW and NCN, Poland; FCT, Portugal; MNE/IFA, Romania; MES of Russia and NRC KI, Russian Federation; JINR; MESTD, Serbia; MSSR, Slovakia; ARRS and MIZ\v{S}, Slovenia; DST/NRF, South Africa; MINECO, Spain; SRC and Wallenberg Foundation, Sweden; SERI, SNSF and Cantons of Bern and Geneva, Switzerland; MOST, Taiwan; TAEK, Turkey; STFC, United Kingdom; DOE and NSF, United States of America. In addition, individual groups and members have received support from BCKDF, the Canada Council, CANARIE, CRC, Compute Canada, FQRNT, and the Ontario Innovation Trust, Canada; EPLANET, ERC, ERDF, FP7, Horizon 2020 and Marie Sk{\l}odowska-Curie Actions, European Union; Investissements d'Avenir Labex and Idex, ANR, R{\'e}gion Auvergne and Fondation Partager le Savoir, France; DFG and AvH Foundation, Germany; Herakleitos, Thales and Aristeia programmes co-financed by EU-ESF and the Greek NSRF; BSF, GIF and Minerva, Israel; BRF, Norway; CERCA Programme Generalitat de Catalunya, Generalitat Valenciana, Spain; the Royal Society and Leverhulme Trust, United Kingdom.

The crucial computing support from all WLCG partners is acknowledged gratefully, in particular from CERN, the ATLAS Tier-1 facilities at TRIUMF (Canada), NDGF (Denmark, Norway, Sweden), CC-IN2P3 (France), KIT/GridKA (Germany), INFN-CNAF (Italy), NL-T1 (Netherlands), PIC (Spain), ASGC (Taiwan), RAL (UK) and BNL (USA), the Tier-2 facilities worldwide and large non-WLCG resource providers. Major contributors of computing resources are listed in Ref.~\cite{ATL-GEN-PUB-2016-002}.
%----------------------------------------------

%% file: atlas_authlist.tex
% ATLAS Collaboration author list
% Reference date of SUSY-2016-26 is 2017-12-07
% Author list last updated on date 17-JUL-18
% Data extracted on 20-Sep-2018 for paper reference SUSY-2016-26
% at 8:43am
 
\begin{flushleft}
{\Large The ATLAS Collaboration}

\bigskip

M.~Aaboud$^\textrm{\scriptsize 34d}$,    
G.~Aad$^\textrm{\scriptsize 99}$,    
B.~Abbott$^\textrm{\scriptsize 125}$,    
O.~Abdinov$^\textrm{\scriptsize 13,*}$,    
B.~Abeloos$^\textrm{\scriptsize 129}$,    
S.H.~Abidi$^\textrm{\scriptsize 165}$,    
O.S.~AbouZeid$^\textrm{\scriptsize 143}$,    
N.L.~Abraham$^\textrm{\scriptsize 153}$,    
H.~Abramowicz$^\textrm{\scriptsize 159}$,    
H.~Abreu$^\textrm{\scriptsize 158}$
% ...
\bigskip
\\

% ...
$^{13}$Institute of Physics, Azerbaijan Academy of Sciences, Baku; Azerbaijan.\\
% ...
$^{34}$$^{(a)}$Facult\'e des Sciences Ain Chock, R\'eseau Universitaire de Physique des Hautes Energies - Universit\'e Hassan II, Casablanca;$^{(b)}$Centre National de l'Energie des Sciences Techniques Nucleaires (CNESTEN), Rabat;$^{(c)}$Facult\'e des Sciences Semlalia, Universit\'e Cadi Ayyad, LPHEA-Marrakech;$^{(d)}$Facult\'e des Sciences, Universit\'e Mohamed Premier and LPTPM, Oujda;$^{(e)}$Facult\'e des sciences, Universit\'e Mohammed V, Rabat; Morocco.\\
% ...
$^{99}$CPPM, Aix-Marseille Universit\'e, CNRS/IN2P3, Marseille; France.\\
% ...
$^{125}$Homer L. Dodge Department of Physics and Astronomy, University of Oklahoma, Norman OK; United States of America.\\
% ...
$^{129}$LAL, Universit\'e Paris-Sud, CNRS/IN2P3, Universit\'e Paris-Saclay, Orsay; France.\\
% ...
$^{143}$Santa Cruz Institute for Particle Physics, University of California Santa Cruz, Santa Cruz CA; United States of America.\\
% ...
$^{153}$Department of Physics and Astronomy, University of Sussex, Brighton; United Kingdom.\\
% ...
$^{158}$Department of Physics, Technion, Israel Institute of Technology, Haifa; Israel.\\
$^{159}$Raymond and Beverly Sackler School of Physics and Astronomy, Tel Aviv University, Tel Aviv; Israel.\\
% ...
$^{165}$Department of Physics, University of Toronto, Toronto ON; Canada.\\
% ...
$^{a}$ Also at  Department of Physics, University of Malaya, Kuala Lumpur; Malaysia.\\
$^{b}$ Also at Borough of Manhattan Community College, City University of New York, NY; United States of America.\\
$^{c}$ Also at California State University, East Bay; United States of America.\\
$^{d}$ Also at Centre for High Performance Computing, CSIR Campus, Rosebank, Cape Town; South Africa.\\
$^{e}$ Also at CERN, Geneva; Switzerland.\\
% ...
$^{*}$ Deceased

\end{flushleft}

% Created with Glance <Atlas.Glance@cern.ch>